\documentclass[%
 reprint,
superscriptaddress,
 amsmath,amssymb,
 aps,
prb,
floatfix,
]{revtex4-2}

\usepackage{textgreek}
\usepackage{booktabs}
\usepackage{array}
\usepackage{multirow}
\usepackage{amsmath, amssymb, amsthm}
\usepackage{mathtools}
\usepackage{bbm}
\usepackage{graphicx}
\usepackage{physics}
\usepackage{tensor}
\usepackage{siunitx}
\usepackage[version=4]{mhchem}
\usepackage{tikz}
\usepackage{xcolor}
\usepackage{listings}
\usepackage{autobreak}
\usepackage[colorlinks,unicode]{hyperref} % , linkcolor=black, anchorcolor=black, citecolor=black, urlcolor=black, filecolor=black
\usepackage{xurl}
\usepackage[most]{tcolorbox}

\usepackage{xr}
\usepackage{prettyref}

\newcommand*{\ii}{\mathrm{i}}
\newcommand*{\ee}{\mathrm{e}}

\lstdefinestyle{console}{
    basicstyle=\footnotesize\ttfamily,
    breaklines=true,
    postbreak=\mbox{\textcolor{red}{$\hookrightarrow$}\space}
}

\newrefformat{fig}{Fig.~\ref{#1}}
\newrefformat{tbl}{Table~\ref{#1}}
\newrefformat{appx}{Appendix~\ref{#1}}

\usetikzlibrary{calc}
\tikzset{every picture/.style={line width=0.3pt}} 


\DeclareSIUnit{\angstrom}{\text{\AA}}

\DeclareSIUnit{\au}{a.u.}

\newcommand*{\bfk}{\mathbf{k}}

\begin{document}

\title{Exciton-driven coherent dressing in transient absorption}
\author{Jinyuan Wu}
\affiliation{Department of Materials Science, Yale University, New Haven, CT 06520}
\author{Chang Shu}
\affiliation{Department of Physics, University of Michigan, Ann Arbor, MI 48109}
%\author{Kai Sun}
%\affiliation{Department of Physics, University of Michigan, Ann Arbor, MI 48109}
\author{Yang-hao Chan}
%\email{yanghao@gate.sinica.edu.tw}
\affiliation{Institute of Atomic and Molecular Sciences, Academia Sinica, Taipei 10617, Taiwan}
\author{Diana Y. Qiu}
\email{diana.qiu@yale.edu}
\affiliation{Department of Materials Science, Yale University, New Haven, CT 06520}

\begin{abstract}
    While coherent excitons have recently been observed to dress electronic states in time- and angle-resolved photoemission spectroscopy (tr-ARPES)\cite{pareek2026driving}, exciton-driven dressing in optical spectroscopy remains largely unexplored. This work develops a theoretical and computational approach to exciton-driven dressing effects in transient absorption spectroscopy in the time-dependent adiabatic $GW$ formalism and derives two symmetry- and band topology-protected selections rules for exciton-dressed transient absorption, unique from tr-ARPES.
    We then numerically calculate exciton-dressed transient absorption spectra of monolayer \ce{MoS2} and demonstrate the selection rules and their lifting after symmetry breaking.
\end{abstract}

\maketitle

\section{Introduction}

Driven quantum systems have emerged as a frontier of condensed matter physics, hosting robust nonequilibrium phases with no equilibrium analog, 
%which, despite being far from thermal equilibrium, often demonstrate stable and robust features beyond conventional classifications of phases of matter.
such as Floquet states \cite{oka2009photovoltaic,grushin2014floquet,giovannini2020floquet,rudner2020floquet}, time crystals \cite{sacha2017time,zaletel2023colloquium}, and active matter states \cite{ramaswamy2010mechanics,marchetti2013hydrodynamics}.
Among these, Floquet states have attracted attention particularly for their experimental accessibility and rich phenomenology.
Floquet systems provide a versatile platform for band engineering by tuning electronic structure through the power and frequency of the driving field, enabling exotic band structures not easily accessible in static systems, including renormalization of the Dirac cones in graphene \cite{merboldt2025observation}, Hall effects without magnetic fields \cite{oka2009photovoltaic}, Floquet-induced fractional Chern insulators \cite{grushin2014floquet}, and creation of Floquet-Weyl semimetals \cite{hubener2017creating}. 

Floquet states form as a consequence of periodic external driving, which is typically but not necessarily from  external optical pumping.
The external driving can be emulated by non-optical bosonic fields within the system itself.
In this case, the coherent oscillation is typically still launched by a strong laser pulse, but the resulting periodic internal driving persists long after the laser pulse ends \cite{giovannini2020floquet}.
Coherent phonons, which have been observed experimentally 
\cite{trovatello2020strongly,he2016coherent,drescher2025onset},
have been theoretically proposed as a possible source for Floquet dressed states \cite{shin2018phonon,giovannini2020floquet,hübener2018phonon}. 
Coherent periodic driving can also originate from magnons \cite{heins2026self}.

Coherent exciton dressing has recently gained attention as another source of periodic driving internal to the material\cite{chan2023giant,pareek2026driving}.
%Excitons are strongly bound electron-hole pairs commonly found in insulators, and their bosonic nature make them another candidate for long-lasting oscillations and hence a possible origin of Floquet dressing.
Theoretical work has shown that coherent excitonic polarization can periodically drive the electronic states, producing Floquet-like modifications of the quasiparticle spectral function
\cite{chan2023giant}. The predicted exciton-driven dressing effects have since been experimentally observed using time- and angle-resolved photoemission spectroscopy (TR-ARPES)\cite{pareek2026driving}.

The observation of coherent exciton dressing effects in TR-ARPES naturally raises the question of whether similar coherent effects can also be observed in optical processes such as transient absorption \cite{kobayashi2023floquet}.
In TR-ARPES, the quasiparticle band structure is probed, and the coherent excitonic polarization field can be intuitively understood as an external driving field remixing the single-quasiparticle states.
In transient absorption dressed by coherent excitons, however, both the pump- and the probe-induced responses are excitonic polarizations, and the coupling between the two resembles a nonlinear optical process rather than Floquet renormalization of fermionic quasiparticles.
This difference in the mechanism of coherent dressing in TR-ARPES and transient absorption may lead to distinct physical consequences, including selection rules in transient absorption that are not present in the same form in TR-ARPES.

This work explores coherent excitonic dressing effects in transient absorption spectroscopy in a first-principles non-equilibrium numerical approach.
In \prettyref{sec:tdagw-arpes-abs}, we introduce the non-equilibrium Green's function formalism, time-dependent adiabatic $GW$ (TD-aGW), and show that it is equivalent to treating coherent excitonic polarization as a nonlinear classical field,
and compare TR-ARPES and transient absorption within this framework.
In \prettyref{sec:no-circular-polarization-renormalization} and \prettyref{sec:no-a-to-a-renormalization}, we prove, respectively, a macroscopic selection rule for coherent exciton dressing in transient absorption based purely on symmetry and a microscopic selection rule arising from band topology.
The exciton-dressed transient absorption spectrum is then numerically simulated for monolayer \ce{MoS2} in \prettyref{sec:compute}, verifying both selection rules.  Under coherent A-exciton driving, first-order oscillations are forbidden in circularly polarized transient absorption spectra, and the absence of first-order self-dressing leads to a second-order response, manifesting as frequency doubling of the transient-absorption oscillations. In contrast, coherent C excitons produce a first-order dressing of the A exciton, with oscillations at the C-exciton frequency.
We conclude in \prettyref{sec:conclusion}.

\section{Distinctions between exciton self-dressing in photoemission and transient absorption}\label{sec:tdagw-arpes-abs}

\begin{figure*}
    \centering
    \includegraphics[width=0.95\linewidth]{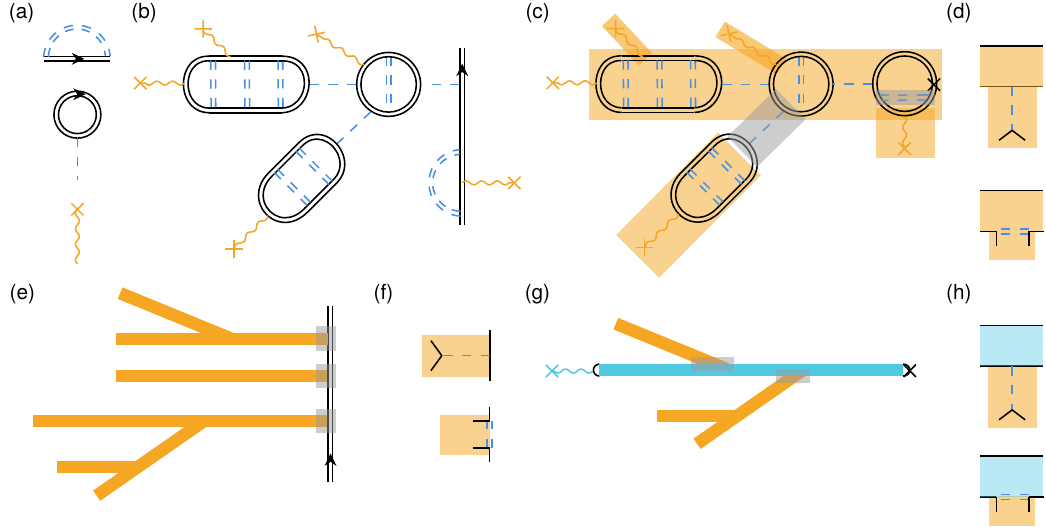}
    \caption{Diagrammatic analysis of TD-aGW. (a) The three parts of the non-equilibrium self-energy in Eq. \eqref{eq:tdagw} from top to bottom: the screened Fock diagram, the Hartree diagram, and the light-matter coupling term. The double lines represent the non-equilibrium Green's function. The blue dashed(solid) lines represent screened(unscreened) Coulomb interactions, and the wiggly orange lines represent external photons.
    (b) An example of self-energy corrections to the non-equilibrium single-quasiparticle Green's function, constructed by recursively applying self-energy corrections in (a). 
    The double lines represent the electron propagator renormalized by the equilibrium $GW$ energy only.
    (c) An illustration of the contribution of (b) to a single-body operator, represented by a black cross; the cross may denote either the dipole operator (relevant in optical simulations; see panel (g)) or one end of the Coulomb interaction line (a part of a higher order diagram renormalizing the single-quasiparticle Green's function).
    The diagram can be seen as a tree, each branch of which either is an external field line or an excitonic ladder-diagrams in equilibrium $GW$-BSE
    (shaded orange).
    The vertices at which these exciton diagrams join (shaded gray), i.e. the exciton-exciton interaction vertex, is also from the $GW$ self-energy, or equivalently, the $GW$-BSE kernel, shown in (d).
    (d) Two possible ways for two exciton diagrams to join, corresponding to the exchange (lower) and direct (upper) parts of the $GW$-BSE kernel.
    (e) Non-equilibrium renormalization of the single-electron propagator can be seen as interaction between the electron and a background classical non-linear excitonic polarization field, which determines exciton-dressed Floquet effects in pumped ARPES.
    Excitonic polarization is represented by the orange bars, the internal structure of which is displayed in (c).
    (f) Interaction between excitonic polarization and the single-electron propagator, from the $GW$-self energy.
    (g) Non-equilibrium excitonic renormalization effects in pump-probe measurements.
    The blue bar denotes probe-induced excitonic polarization, which is assumed to be linear and has the usual $GW$-BSE ladder diagram internal structure.
    As in (c), the probe-induced excitonic polarization diagram is corrected by pump-induced excitonic polarization (in orange). (h) illustrates the interaction vertices between pump- and probe-induced excitonic diagrams.
    }
    \label{fig:tdagw}
\end{figure*}

\subsection{Single-particle response in time-resolved photoemission}

Ultrafast spectroscopic techniques are often simulated by non-equilibrium Green's function methodologies.
In this article, non-equilibrium responses are described by generalizing the many-body perturbative $GW$ formalism \cite{hedin1965new,hedin1970effects,hybertsen1986electron,adler1962quantum,wiser1963dielectric} to non-equilibrium, where we make an adiabatic approximation by ignoring retardation effects and replacing the full GW self energy by the static COHSEX approximation \cite{hybertsen1986electron}, which reduces the formalism into a quantum master equation known as time-dependent adiabatic $GW$ (TD-aGW) \cite{attaccalite2011real,perfetto2015nonequilibrium,sangalli2021excitons,chan2021giant,chan2023giant}
\begin{equation}
    \ii\frac{\partial}{\partial t}\rho(t) = \left[ H^\text{QP}  + \delta\Sigma^{GW}(t) - e\textbf{E}(t)\cdot\textbf{r}, \rho(t) \right],
    \label{eq:tdagw}
\end{equation}
where $\rho(t) = \rho_{nm\textbf{k}}(t)$ is the single-electron reduced density matrix,
$m, n$ are the band indices of the density matrix and $\vb{k}$ the momentum of the electron,
$\var{\Sigma^{GW}}$ is the non-equilibrium variation of $GW$ energy (the equilibrium part of which has been integrated into $H^{\text{QP}}$ and calculted using equilibrium $GW$), approximated by the static COHSEX self-energy.
If retardation and nonequilibrium changes in screening are neglected, functional derivative of the $GW$ self energy yields the kernel of the $GW$-Bethe-Salpeter equation ($GW$-BSE) \cite{rohlfing2000electron}, allowing
$\var{\Sigma^{GW}}$ to be written as \cite{attaccalite2011real,perfetto2015nonequilibrium,hou2025data}
\begin{equation}
    \var{\Sigma}^{GW}_{nm\vb{k}} = \sum_{n'm'\vb{k}'} K_{nm\vb{k} n'm'\vb{k}'} \var{\rho}_{n'm'\vb{k}'},
\end{equation}
where $\var{\rho}$ is the time-dependent change in the single-electron reduced density matrix 
%$ = \rho - \rho^{\text{eq}}$
and $K_{nm\vb{k}n'm'\vb{k}'}$ is the $GW$-BSE kernel \cite{hou2025data}.
The linear response \eqref{eq:tdagw} is equivalent to equilibrium linear absorption spectroscopy calculated using $GW$-BSE
\cite{attaccalite2011real,chan2021giant,hou2025data}.

TD-aGW has been used to study coherent excitonic dressing effects in ARPES~\cite{chan2023giant,stefanucci2026unified}. To understand how this physics emerges within the formalism, we first examine the underlying diagrammatic processes captured by TD-aGW (\prettyref{fig:tdagw}). Because TR-ARPES probes the single-quasiparticle electronic structure, the relevant diagrams renormalize the single-electron Green’s function.
A representative diagram is shown in \prettyref{fig:tdagw}(b).
The renormalization arises from a series of non-equilibrium Hartree and Fock diagrams, which in turn are renormalized by the same diagrammatic components and ultimately terminate at external field lines.

These diagrams can be reorganized into a complementary picture in which the quasiparticle interacts with a coherent microscopic polarization field (\prettyref{fig:tdagw}(e)).
A diagrammatic representation of the microscopic electromagnetic polarization density in TD-aGW is given in (\prettyref{fig:tdagw}(c)).
By diagrammatic manipulation, it can be seen that each Feynman diagram contributing to polarization can be recast into a tree structure, with each branch being a $GW$-BSE ladder diagram.
The polarization density thus can be understood to have an excitonic origin.
The branches are joined together through nonlinear couplings generated by
%by nonlinearity of this excitonic polarization field, which is itself from 
the $GW$-BSE kernel (\prettyref{fig:tdagw}(d)).
This tree structure of non-equilibrium exciton diagrams means that polarization calculated within TD-aGW is equivalent to treating excitonic polarization as a classical field \cite{boulware1968tree}.
The non-equilibrium Fock term is usually not localized in time,
but after the static COHSEX approximation, retardation effects are removed.
As a result, the excitonic polarization field does not have double time dependence (\prettyref{appx:td-gw}), and thus further resembles a classical field.
Therefore, the total polarization in the system, when calculated within the TD-aGW approach, contains and \emph{only} contains exciton self-driving effects in which both the single exciton propagator and exciton-exciton interaction is described by the $GW$-BSE kernel (\prettyref{fig:tdagw}(c, d)).
This can be seen as a nonlinear generalization of the equivalence between linear TD-aGW and $GW$-BSE.

Thus, all non-equilibrium renormalization of the single-electron Green's function 
%, the equal time part of which is $\rho$ in TD-aGW,
can be seen as the electron interacting with the nonlinear excitonic polarization field one or more times, with the corresponding interaction vertices again determined by the $GW$-BSE kernel (\prettyref{fig:tdagw}(e)).
This is consistent with the picture provided in Ref.~\cite{chan2023giant}.
Thus, TD-aGW calculation of the single-electron Green's function captures and only captures coherent excitonic dressing of the single-electron propagator through its coupling to the nonlinear excitonic polarization field described by a generalization of $GW$-BSE.
When the excitonic polarization field is periodic and is strong enough, coherent dressing leads to Floquet effects, producing band replicas, avoided crossings, and camelbacking \cite{chan2023giant,perfetto2019pump}.

\subsection{Excitonic response in transient absorption}\label{sec:method}

TD-aGW can also be utilized to simulate transient absorption experiments.
In typical transient absorption spectroscopic experiments, a strong (but commonly low-frequency) pump pulse arrives first, exciting the system into non-equilibrium, whose state is then characterized by the energy loss of a much weaker probe signal going through the system, with a delay time of $\tau$ and a central frequency of $\omega$.
Following Refs.~\cite{perfetto2015some,perfetto2015nonequilibrium}, in monolayer materials in non-equilibrium, when the pump is off, the transient absorption spectrum can be defined as (\prettyref{appx:theory-absorption-detail})
\begin{equation}
    \text{absp. at $\omega, \tau$} \sim \Im \frac{\vu{e}^* \cdot \vb{d}(\omega)}{{E}_\tau(\omega)} \sim \Im \vu{e}^\dagger \cdot \vb{\chi_\tau}(\omega) \cdot \vu{e} ,
    \label{eq:absorption-2d}
\end{equation}
where the total electric field in the system is $\vb{{E}}(t) = \vu{e} E_\tau(t)$, and $E_\tau(t) = E(t - \tau)$, $E(t)$ is the temporal profile of the wave packet, 
and $\chi_\tau(\omega)$ is the electric susceptibility tensor of the pumped system to the probe $E_\tau(t)$.
Because of its mathematical form, Eq.~\eqref{eq:absorption-2d} may be known as the non-equilibrium generalization of the imaginary part of the macroscopic dielectric function \cite{sato2014numerical},
although in two-dimensional systems, strictly speaking, the macroscopic dielectric function cannot be well-defined and interpreting Eq.~\eqref{eq:absorption-2d} as the macroscopic dielectric function is only based on shared mathematical structure, without clear physical motivations \cite{liu2025direct}.
Within the TD-aGW framework, the transient absorption spectrum (Eq.~\eqref{eq:absorption-2d}) is obtained by collecting the expectation value of the dipole at each time step, taking its Fourier transform, dividing by the frequency domain profile of the probe pulse, and taking the imaginary part.
%To improve the computational efficiency, we utilize the fact that the response of the pumped system to the probe is mostly linear,
%and replace frequency sweeping in Eq.~\eqref{eq:absorption-2d} by replacing the physically realistic probe pulse by a very narrow gaussian wave packet,
%which includes a frequency bandwidth that is wide enough to accurately calculate Eq.~\eqref{eq:absorption-2d}.
Because the response to the weak probe is approximately linear, the frequency-dependent transient absorption spectrum can be obtained from a single time propagation rather than by sweeping the probe frequency. We therefore use a temporally narrow Gaussian probe, whose broad frequency bandwidth spans the spectral range of interest, and extract the frequency-dependent response from the Fourier transform of the induced dipole.
The actual experimentally observed transient absorption spectra are further subject to broadening due to the frequency-domain spectral width of the probe wave packet.

An important coherent phenomenon in Eq.~\eqref{eq:absorption-2d} is temporal oscillation of the absorption spectrum, because $\vb{d}(\omega)$ depends on the temporal profile of $\vb{E}$ and hence the delay time between the pump and the probe.
When the system undergoes periodic driving,
we expect the transient absorption spectrum \eqref{eq:absorption-2d} to oscillate periodically as $\tau$ varies,
with the same period of the driving field \cite{rudner2020floquet}, and 
\begin{equation}
    \text{absp. at $\omega, \tau$} \sim \Im \sum_n \chi^{(n)}(\omega) \frac{E(\omega - n \omega_0)}{E(\omega)},
    \label{eq:absorption-2d-omega}
\end{equation}
where $\chi^{(n)}$ is the Floquet frequency-domain dielectric response, $E(\omega - n \omega_0) \sim \ee^{\ii (\omega - n \omega_0) \tau}$, and $\omega_0$ is the driving field frequency.
Such oscillations are observed with coherent phonons \cite{trovatello2020strongly,he2016coherent,drescher2025onset} but are absent in optically dressed transient absorption experiments \cite{kobayashi2023floquet}.
The appearance of such oscillations in numerical simulation is cited by Ref.~\cite{sato2014numerical} to argue that the imaginary part of the response of the polarization to the total electric field has no physical meaning in non-equilibrium systems.
Their absence in experiment, however, may simply reflect limitations of measurements.
Typical transient-absorption measurements have a temporal resolution above $\sim \SI{10}{fs}$,
whereas the lowest Floquet driving frequency in Ref.~\cite{kobayashi2023floquet}, $\sim \SI{0.18}{eV}$, corresponds to a period of \SI{23}{fs}, which is only marginally above the Nyquist period limit. 
The oscillation signals may further be blurred away by the random carrier-envelope phase.
The delay-dependent $n\neq0$ components are therefor averaged out, leaving only $\Im \chi^{(n=0)}(\omega)$, which corresponds to the temporally averaged transient absorption spectrum.
When the Floquet quasi-stationary states are available, this quantity can be evaluated using the same Fermi golden rule expression typically used to simulate equilibrium absorption spectroscopy \cite{gatti2013exciton}
by treating the Floquet quasi-stationary states as if they were ordinary wave functions \cite{rudner2020floquet}.
In particular, the frequencies of the peaks in the transient absorption spectrum are identical to the Floquet quasi-eigenvalues.
This indicates that Floquet renormalization of exciton states is reflected by peak shifts in the experimentally observed transient absorption spectrum in most experimental setups.

Incoherent exciton populations can also produce spectral changes that are qualitatively similar to the time-averaged effects of coherent exciton-driven dressing, for example through Pauli blocking. A numerical comparison between the time-averaged ($n=0$) coherent response and the response of a dephased nonequilibrium state is presented in \prettyref{fig:tr-abs}.
Similar ambiguity arises in TR-ARPES, where qualitatively similar spectral renormalization can arise from both coherent and incoherent exciton populations \cite{stefanucci2026unified}.
Thus, delay-dependent oscillations of the spectrum provide a more robust indication of coherent exciton-driven dressing than time-averaged spectral changes alone.

A diagrammatic analysis of transient absorption in TD-aGW is given in \prettyref{fig:tdagw}(g).
\prettyref{fig:tdagw}(c) can be seen as a representative diagram of the total macroscopic polarization in the system, which can be calculated by contraction of the single-electron density matrix and the dipole matrix. 
In the linear probe regime, in each diagram, there should be only one external line corresponding to the probe, which is connected to the dipole matrix by a series of ladder diagrams renormalized by other non-linear polarization diagram.
Therefore, after reorganizing the diagrams, linear absorption on top of a background of coherent excitons can be seen as planting zero, one or more excitonic polarization diagram corrections to each propagator in a standard equilibrium $GW$-BSE diagram (\prettyref{fig:tdagw}(g)).
From the perspective of the quantum master equation formulation of TD-aGW, in a pump-probe calculation, the non-equilibrium density matrix can be decomposed into $\var{\rho} = \var{\rho^{\text{pump}}} + \var{\rho^{\text{probe}}}$, 
where $\var{\rho^{\text{pump}}}$ alone satisfies the equation of motion, Eq. \eqref{eq:tdagw}, in which $\vb{E}$ only contains the pump electric field,
and $\var{\rho^{\text{probe}}}$ is the density matrix change on top of the pump-induced density matrix.
The non-equilibrium self-energy renormalization can similarly be decomposed into $\var{\Sigma^{GW}} = \var{\Sigma^{\text{pump}}} + \var{\Sigma^{\text{probe}}}$,
where $\var{\Sigma^{\text{pump}}} = K \var{\rho^{\text{pump}}}$ and $\var{\Sigma^{\text{probe}}} = K \var{\rho^{\text{probe}}}$.
Ignoring terms non-linear with respect to the probe field, the equation of motion of $\var{\rho^{\text{probe}}}$ is given by 
\begin{widetext}
    \begin{equation}
    \begin{aligned}
        \ii \pdv{t} \var{\rho^{\text{probe}}} &=  [H^{\text{QP}}, \var{\rho^{\text{probe}}}]  + [- e \vb{E}^{\text{probe}}(t) \cdot \vb{r} + \var{\Sigma}^{\text{probe}}, \rho^{\text{eq}}] \\
        &+ \underbrace{[\var{\Sigma^{\text{probe}}}, \var{\rho^{\text{pump}}}] + [\var{\Sigma^{\text{pump}}}, \var{\rho^{\text{probe}}}]  + [-e \vb{E}^{\text{probe}} (t) \cdot \vb{r}, \var{\rho^{\text{pump}}}]}_{\text{self-driving field}} + [-e \vb{E}^{\text{pump}} (t) \cdot \vb{r}, \var{\rho^{\text{probe}}}].
    \end{aligned}
    \label{eq:probe-eom}
\end{equation}
\end{widetext}
The first two terms correspond to linear BSE \cite{attaccalite2011real}, and the remaining terms can be understood as the self-driving influence to the structure of excitons excited by the probe.
It can be verified that a Feynman diagrammatic representation of Eq.~\eqref{eq:probe-eom} in which the pump and the probe are treated as different external driving fields is identical to \prettyref{fig:tdagw}(g).
We note that the last $[-e \vb{E}^{\text{pump}} (t) \cdot \vb{r}, \var{\rho^{\text{probe}}}]$ term vanishes in this study, as the probe starts after the pump ends.

Importantly, TR-ARPES is not described by Eq.~\eqref{eq:probe-eom} because the probe photoemits an electron from the material rather than inducing an additional polarization within it. The ARPES spectrum is instead constructed analytically using the two-time single-electron Green's function in \prettyref{fig:tdagw}(e), and the dynamics of the photoemission process itself are not modeled within TD-aGW \cite{chan2023giant}. 
Comparing \prettyref{fig:tdagw}(e) and (g), we see that in TR-ARPES, the coherent exciton field couples two single-particle states, while in transient absorption, the coherent exciton field couples two exciton states.
We therefore expect TR-ARPES and transient absorption to obey different slection rules imposed by the macroscopic symmetry (\prettyref{sec:no-circular-polarization-renormalization}).
Furthermore, from Eq.~\eqref{eq:probe-eom}, it is clear that in transient absorption, the dressed probe polarization is additionally projected onto coherent optical excitons ($\sim \var{\rho^{\text{pump}}}$).
This extra contraction over electron-hole amplitudes can make a symmetry-allowed coupling cancel over momentum, giving rise to additional microscopic selection rules absent in TR-ARPES (\prettyref{sec:no-a-to-a-renormalization}).

\section{Selection rules for exciton self-dressing in transient absorption}\label{sec:selection-rule}

\subsection{A macroscopic polarization selection rule}\label{sec:no-circular-polarization-renormalization}

In this section, we prove that in a two-dimensional material with $C_3$ rotational symmetry, coherent excitonic polarization has no first-order contribution to the temporal oscillation of circularly polarized transient absorption spectra.
We note that this selection rule is highly generic and applies to many two-dimensional transition-metal dichalcogenides (TMDs).
Further, we show that in materials with other types of rotational symmetries, coherent excitonic polarization has no first-order contribution to transient absorption spectra at all, regardless of polarization. 

All complex irreducible representations of the in-plane rotation group $SO(2)$ are one-dimensional, and are labeled by the so-called two-dimensional angular momentum quantum number $m$:
when the system is rotated by an angle $\theta$, a physical quantity symmetric under two-dimensional rotation and carrying an irreducible representation of $SO(2)$ acquires a phase factor $\ee^{\ii m \theta}$ under this rotation.
The two-dimensional vector space, containing all possible electric polarization directions, can be linearly decomposed into left- and right-circular polarization, which have $m = + 1$ and $m=-1$, respectively, and linearly polarized light is a superposition of these two components.
Similarly, in the two-dimensional space, the electric susceptibility tensor $\chi_{ij} = \frac{N}{\epsilon_0} \pdv{d_i}{E_j}$ is a rank-two tensor and can be decomposed into the irreducible representations of $SO(2)$ as $0 \oplus 0 \oplus 2 \oplus -2$.
The four components are $\chi_{xx} + \chi_{yy}$ ($m=0$), $\chi_{xy} - \chi_{yx}$ ($m=0$), $\chi_{xx} - \chi_{yy} + \ii (\chi_{xy} + \chi_{yx})$ ($m=+2$), and $\chi_{xx} - \chi_{yy} - \ii (\chi_{xy} + \chi_{yx})$ ($m=-2$).

The electric polarization direction $\vu{e}$ in linear polarization has the form of $(\cos \theta, \sin \theta)$, and thus only $\chi_{xx} + \chi_{yy}$ and the $m = \pm 2$ components contribute to Eq.~\eqref{eq:absorption-2d}.
On the other hand, in circularly polarized probing, $\vu{e} = \frac{1}{\sqrt{2}} \pmqty{1 \\ \pm \ii}$, so the transient absorption spectrum in Eq.~\eqref{eq:absorption-2d} contains only the $m=0$ terms.
These observations also follow from angular momentum conservation, as in circularly polarized absorption spectroscopy, the angular momentum of the photon is conserved, while linearly polarized absorption spectroscopy involves interaction channels flipping the photon angular momentum. 

In systems with $C_3$ symmetry, the angular momentum quantum number $m$ is conserved modulo 3 \cite{zhang2018optical}.
Because coherent exciton polarizations in a system ultimately come from optical pumping, which only has $m=\pm 1$ components, the first-order effects of the coherent exciton polarization in $\var{\rho}^{\text{pump}}$ felt by $\chi$ either have $m=\pm 1$ or $m=\pm 2$ characteristics, but not $m=0$ characteristics.
As such, the $m=0$ components of the tensor $\chi$ has no linear response to the coherent exciton polarization, which oscillates in the frequency of the coherent exciton.
From this it immediately follows that no first-order influence of the coherent exciton population exists in the the circularly polarized transient absorption spectra.
On the other hand, the first-order non-equilibrium effect in linearly polarized transient absorption spectrum exists only in the $m = \pm 2$ part.
This means that the first-order non-equilibrium renormalization of the absorption spectrum can be summarized as a rotating preferred absorption axis, as is numerically shown in \prettyref{fig:oscillate} in \prettyref{appx:c-drive}.
Therefore, an angle-averaged absorption spectrum contains no first-order effects of the coherent exciton polarization, but a polarization-resolved linear probe can detect an oscillating optical anisotropy.
This selection rule of absence of first-order $m=0$ renormalization can be lifted by straining the system and removing the $C_3$ symmetry, as numerically demonstrated in \prettyref{fig:strain}.

The selection rule is a generalization of the selection rule in Ref.~\cite{tang1971selection} to scenarios of internal driving: the first-order non-equilibrium renormalization of Eq.~\eqref{eq:absorption-2d} can be seen as a three-photon interaction vertex with one photon line replaced by an exciton line. 
We note that the reasoning above can be generalized by systems with other rotational symmetric groups.
Since none of $m=0,\pm 2$ satisfies the angular momentum conservation relation $m=\pm 1 \pmod n$  when $n=2,4,$ or $6$, we conclude that first-order renormalization of the transient absorption spectrum is forbidden for systems with $C_n$ symmetry where $n \neq 3$, regardless of polarization. 

\subsection{A microscopic band-winding selection rule}\label{sec:no-a-to-a-renormalization}

The selection rule in \prettyref{sec:no-circular-polarization-renormalization} relies solely on the macroscopic symmetry and group representation.
In this section, we prove a microscopic, band topology-induced selection rule that disallows symmetry-allowed non-equilibrium renormalization:
we assert that an optically active two-band exciton confined in a topologically non-trivial valley of a two-dimensional material with $C_n$ ($n \geq 3$) symmetry but without inversion symmetry cannot have first-order coherent driving of itself.

We start by writing down a two-band effective model of the valley.
A $\vb{k}\cdot \vb{p}$ model of a valley with the $C_n$ symmetry in the band structure of a two-dimensional material takes the generic form of \cite{zhang2018optical} (\prettyref{appx:kphamiltonian})
\begin{equation}
    H = d_0(\abs{\vb{k}}) + \pmqty{
    \Delta(\abs{\vb{k}}) & \alpha(\abs{\vb{k}}) \ee^{\ii w \phi_{\vb{k}}} \\ 
    \alpha(\abs{\vb{k}}) \ee^{-\ii w \phi_{\vb{k}}}  & - \Delta(\abs{\vb{k}}) \\
    },
    \label{eq:two-band-ham}
\end{equation}
where the basis is $\ket{u_{c\vb{k}=0}}$ and $\ket{u_{v\vb{k}=0}}$, $2\Delta(\vb{k}=0)$ is the band gap, and $\phi_{\vb{k}}$  is the angle of $\vb{k}$.
The basis is $\ket{u_{c\vb{k}=0}} \approx \ket{d_{z^2}}$ and $\ket{u_{v\vb{k}=0}} \approx (\ket{d_{x^2-y^2}} \pm \ii \ket{d_{xy}}) / \sqrt{2}$ for monolayer \ce{MoS2} \cite{xiao2012coupled}, but the selection rule to be proven does not rely on the specific atomic orbitals contained in the states.
For convenience, we have shifted the origin of the momentum space coordinates to the center of the valley.
The amplitude of the non-diagonal Hamiltonian matrix element, $\alpha$, vanishes at $\vb{k} = 0$.
The integer $w$ controlling the winding of the non-diagonal terms of the Hamiltonian around the valley and is thus known as the winding number.
Its significance can be understood by diagonalizing the Hamiltonian \eqref{eq:two-band-ham}.
We find that the eigenstates of the Hamiltonian are
\begin{equation}
    \ket{u_{c \vb{k}}} = \pmqty{
        \cos \frac{\theta_{\vb{k}}}{2} \\
        \ee^{-\ii w \phi_{\vb{k}}} \sin \frac{\theta_{\vb{k}}}{2}
    }, \quad 
    \ket{u_{v \vb{k}}} = \pmqty{
        - \ee^{\ii w \phi_{\vb{k}}} \sin \frac{\theta_{\vb{k}}}{2} \\
        \cos \frac{\theta_{\vb{k}}}{2}
    },
    \label{eq:eigenstates-mos2}
\end{equation}
and the band energies are $E_{c,v\vb{k}} = \pm E_{\vb{k}} = \pm \sqrt{\Delta^2 + \abs{\alpha}^2}$,
and $\cos \theta_{\vb{k}} = \Delta / E_{\vb{k}}$.
We note that $\theta_{\vb{k}}$ vanishes when $\vb{k}=0$ where $\alpha$ vanishes.
It now appears that $w$ gives the winding of the relative phases between the two components of the wave function, and a non-zero $w$ means that the relative phase between the two components goes around the origin $w$ times when $\vb{k}$ goes around the origin once.
This topologically non-trivial winding behavior means that a non-zero Berry curvature occurs near the valley, and in this case the valley is topologically nontrivial.
Assuming time-reversal symmetry, this is only possible when the (three-dimensional) inversion symmetry is broken and a non-zero local Berry curvature is allowed \cite{zhang2018optical,cayssol2021topological}.
For the specific case of monolayer \ce{MoS2} considered below, the $K$ and $K'$ valleys have winding numbers $w=\pm 1$ \cite{xiao2012coupled,cao2012valley}. We emphasize that the winding number $w$ describes the topological features of the valley and is not a single-quasiparticle quantum number from the symmetry representation it carries, unlike the angular momentum quantum number $m$ introduced in \prettyref{sec:no-circular-polarization-renormalization}.

Now we consider optical transition in the valley.
The oscillator strength of an exciton mode $S$ is determined by contraction between the exciton envelope wave function in the Bloch basis $A^S_{vc\vb{k}}$ and the dipole matrix element in the Bloch basis $\mel{v\vb{k}}{\vb{r}}{c\vb{k}}$.
In a valley with $C_n$ symmetry, $\ket{u_{c\vb{k}=0}}$ and $\ket{u_{v\vb{k}=0}}$ transform according to irreducible representations of $C_n$, and hence have well-defined in-plane  angular momenta $m_c$ and $m_v$ modulo $n$ \cite{zhang2018optical}.
For an interband transition at the valley center to be optically active, the dipole matrix element must be symmetry allowed.
Since the two circular components of the dipole operator carry angular momentum $\Delta m=\pm 1$, this requires
\begin{equation}
    m_c-m_v=\pm 1 \pmod n. \label{eq:band-angular-momentum-dipole-rule}
\end{equation}
For monolayer \ce{MoS2}, for example,
$m_c-m_v=\pm 2\equiv\mp 1\pmod 3$, consistent with the optical activity of the $K$ and $K'$ valleys.
In contrast, the corresponding matrix elements between identical band-edge states vanish by symmetry, since these states have $\Delta m=0$.

Near the valley center, $\cos(\theta_{\vb{k}}/2)\approx 1$, and the $\sin (\theta_{\vb{k}} / 2)$ terms in the dipole matrix elements vanish due to the angular momentum conservation rule Eq.~\eqref{eq:band-angular-momentum-dipole-rule}. Therefore,
\begin{equation}
    \mel{v\vb{k}}{\vb{r}}{c\vb{k}}
    \approx
    \mel{u_{v\vb{k}=0}}{\vb{r}}{u_{c\vb{k}=0}},
    \label{eq:dipole-sign-structure}
\end{equation}
Thus, to leading order, the interband dipole has no angular winding around the valley in the gauge of Eq.~\eqref{eq:eigenstates-mos2} (\prettyref{fig:no-first-order}(c)).
Because the exciton amplitude $A^S_{vc\vb{k}}$ has a well-defined envelope angular momentum modulo $n$ (\prettyref{appx:symmetry-gw-bse}), an optically active exciton must therefore have zero envelope angular momentum modulo $n$ in this gauge.

We then turn to evaluate the self-driving field in Eq.~\eqref{eq:probe-eom} under the first-order approximation, in which we assume that $\var{\rho}^{\text{pump}}_{cv\vb{k}} \propto A^S_{vc\vb{k}} \ee^{-\ii \Omega^S t}$.
This ignores saturation effects and exciton occupations on other modes due to nonlinearity.
Because the probe is linear, without loss of generality we assume $\var{\rho}^{\text{probe}}_{cv\vb{k}} \propto A^S_{vc\vb{k}} \ee^{-\ii \Omega^S t}$.
We note that there is a phase difference between $\var{\rho^{\text{pump}}}$ and $\var{\rho^{\text{probe}}}$ determined by the delay between the pump and the probe.
Treating Eq.~\eqref{eq:probe-eom} as the non-equilibrium generalization of linear BSE, the self-driving effects from an $S$ exciton to itself can be singled out by projecting $\var{\rho^{\text{probe}}_{vc\vb{k}}}$ to $A^S_{vc\vb{k}}$.
A straightforward calculation shows that because both the dipole matrix and $\var{\rho^{\text{pump}}}$ under the first-order approximation are purely non-diagonal with respect to band indices, $\comm{\vb{r}}{\var{\rho^{\text{pump}}}}$ is purely diagonal, and vanishes once projected to $A^S_{vc\vb{k}}$.
The first-order self-driving effects therefore solely come from $\comm{\var{\Sigma}^{\text{pump}}}{\var{\rho^{\text{probe}}}} + \comm{\var{\Sigma^{\text{probe}}}}{\var{\rho^{\text{pump}}}}$.
Because we intend to study the dependence of the transient absorption spectrum on the delay time, and hence the dependence of the self-driving field on the delay time, without loss of generality, we calculate the self-driving field at a time $t$ when the phase factor of $\var{\rho^{\text{probe}}}$ is unity; $t$ can thus be seen as the delay between the pump and the probe with a constant shift. 
In our two-band scenario, the commutators after projection to $A^S_{vc\vb{k}}$ are
\begin{equation}
    \begin{aligned} 
        &\quad \sum_{\vb{k}} A^{S*}_{vc\vb{k}} (\comm{\var{\Sigma}^{\text{pump}}}{\var{\rho^{\text{probe}}}} + \comm*{\var{\Sigma^{\text{probe}}}}{\var{\rho^{\text{pump}}}})_{cv\vb{k}} \\
        &= 2\sum_{\vb{k} \vb{k}'} \abs*{A_{v c \vb{k}}^{S}}^2 (K_{cc\vb{k} c v \vb{k}'} - K_{vv\vb{k} c v \vb{k}'}) A_{v c \vb{k}'}^{S} \ee^{-\ii \Omega^S t} \\ 
        &+ 2\sum_{\vb{k}\vb{k}'} \abs*{A_{v c \vb{k}}^{S}}^2 (K_{cc\vb{k} v c \vb{k}'} - K_{vv\vb{k} v c \vb{k}'}) A_{v c \vb{k}'}^{S*} \cos(\Omega^S t).
    \end{aligned}
    \label{eq:a-to-a-driving}
\end{equation}
Therefore, whether exciton $S$ has first-order self-dressing reduces to the question of the structure of $K_{vv\vb{k} cv\vb{k}'}$.
Because the bare and screened Coulomb interaction potential is invariant under $C_n$ rotation, and the Bloch states at the center of the valley all carry one-dimensional representations of the $C_n$ rotation group which satisfy the constraint Eq.~\eqref{eq:band-angular-momentum-dipole-rule} for the valley to be optically active, the matrix elements $K_{cc\vb{k}cv\vb{k}'}$ and $K_{cc\vb{k} cv\vb{k}'}$, when perceived as matrices acting on the excitonic Hilbert space, either increase or decrease the exciton envelope angular momentum quantum number by $w$ (\prettyref{appx:symmetry-gw-bse}).
The envelope angular momentum quantum numbers of $\abs{A^S_{vc\vb{k}}}^2$ and $A^{S*}_{vc\vb{k}'}$ are both zero modulo $n$, leading to a vanishing first-order self-driving field in Eq.~\eqref{eq:a-to-a-driving}.
We have therefore shown that, for an optically active two-band exciton confined to a topologically nontrivial valley with $C_n$ symmetry, the first-order coherent self-driving term vanishes.
%The selection rule has now been proven.

More generally, following the derivation above, we are able to show that an optically active exciton mode's coherent population is unable to have first-order dressing effects to another optically active exciton mode:
in this scenario, the $\abs{A}^2$ factor in Eq.~\eqref{eq:a-to-a-driving} is replaced by the multiplication of two exciton wave functions, both of which however carry zero envelope angular momentum modulo $n$, while the $GW$-BSE kernel has a non-zero angular momentum, and the summation vanishes.

We note that the proof is gauge-independent.
It is possible to apply a gauge transform to $\ket{vc\vb{k}}$ that removes the angular momentum in $K_{cc\vb{k} cv\vb{k}'}$.
This however will drive away the phase structure to the dipole, giving it a non-trivial phase factor, which then requires optically active excitons to have a non-zero envelope angular momentum quantum number, and it can be verified that Eq.~\eqref{eq:a-to-a-driving} still vanishes, provided that the valley topological index $w$ is non-zero.

This selection rule is lifted by breaking any of its conditions.
When the pumping exciton is not confined to the valley, the symmetric analysis of the $GW$-BSE kernel in \prettyref{appx:symmetry-gw-bse} fails, and we should in general to expect a non-zero first-order effect.
When the exciton modes involves are multi-band, the self-driving field involves non-trivial contraction of band indices, and the analysis based solely on the angular momentum fails. 
Further, first-order self-driving effects can be recovered by restoring the three-dimensional inversion symmetry, which prohibits non-zero winding numbers in insulators, or by removing the two-dimensional $C_n$ symmetry so the dipole matrix and the $GW$-BSE kernel no longer have block structures under the basis of the conduction band minimum and the valence band maximum, which is a condition necessary to relate the two quantities to the winding number.
This can typically be realized by straining the sample.

\section{First-principles numerical simulations}\label{sec:compute}

We now test the selection rules derived in \prettyref{sec:selection-rule} using first-principles TD-aGW simulations of monolayer \ce{MoS2}.

\subsection{Computational approach}

\begin{figure}
\centering
    \includegraphics[width=\linewidth]{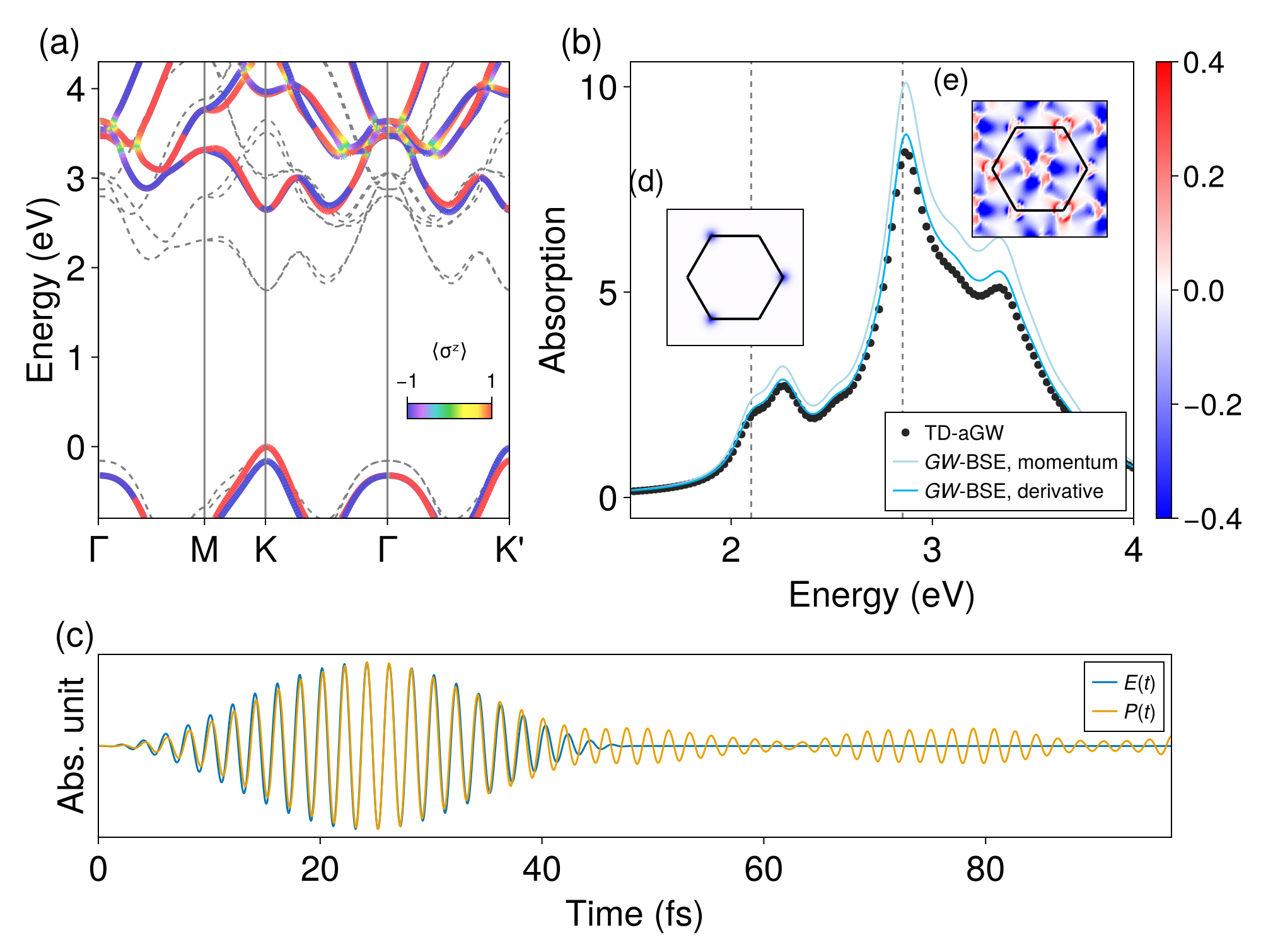}
    \caption{
    (a) The band structure of monolayer \ce{MoS2}, calculated using $G_0W_0$@PBE.
    The gray curves are DFT bands.
    Red and blue colors correspond to spin-up and spin-down states, consistent with previous calculations.
    The lowest spin-up and spin-down conduction bands have a small crossing near the K and K' valleys.
    (b) Benchmark of linear absorption in TD-aGW. The TD-aGW and equilibrium $GW$-BSE calculations are both performed on the $24 \times 24 \times 1$ $\vb{k}$-grid.
    The equilibrium $GW$-BSE absorption spectrum is Lorentzian broadened. 
    The TD-aGW polarization is equivalently regularized in time domain by an exponentially decaying signal before Fourier transform.
    (c) Wave packet envelope of the pump field and excitonic polarization field in the system.
    (d-e) The real part of $\rho_{cv\vb{k}}$ when the A or C exciton is excited, plotted in the first Brillouin zone.
    Here $c$ refers to the lowest spin-up band and $v$ refers to the highest valence bands.
    The single-electron gauge is chosen so that the 1s exciton's wave function is real.
    }
    \label{fig:setup}    
\end{figure}

For TD-aGW, we start by performing a conventional one-shot $G_0W_0$ calculation. The mean-field starting point is obtained from density functional theory (DFT) as implemented in the Quantum ESPRESSO software package\cite{giannozzi2009quantum,giannozzi2017advanced,giannozzi2020quantum,carnimeo2023quantum}, and the GW-BSE calculation is implemented in BerkeleyGW \cite{deslippe2012berkeleygw,shao2016structure,barker2022spinor}.
We start from a monolayer \ce{MoS2} structure with a lattice constant of $a = \SI{3.145}{\angstrom}$ and a distance of $h = \SI{3.143}{\angstrom}$ between the two layers of S atoms. The calculation cell has $\SI{20}{\angstrom}$ of vacuum in the aperiodic direction.
The DFT calculation is performed using the Perdew-Burke-Ernzerhof (PBE) functional \cite{perdew1996generalized} norm-conserving pseudopotentials with a cutoff energy of \SI{95}{Ry}.
The $GW$ calculation is at the $G_0 W_0$ level. The calculation uses a dielectric cutoff energy of \SI{20}{Ry}, a $24 \times 24 \times 1$ $\vb{k}$-grid, and and 6000 bands. Spin-orbit coupling is included within a relativistic spinor formalism~\cite{barker2022spinor}, and Coulomb truncation is used to prevent spurious interactions between periodic images~\cite{ismail2006truncation}.
The band structure is shown in \prettyref{fig:setup}(a), with a $G_0W_0$ band gap of \SI{2.65}{eV}, consistent with previous results \cite{qiu2013optical,qiu2016screening}.

An equilibrium $GW$-BSE calculation is performed on top of the band structure, which produces the $GW$-BSE kernel used in the subsequent TD-aGW calculations.
Because of limitations in computational capacity, the $\vb{k}$-grid of the kernel used in TD-aGW task is $24 \times 24 \times 1$. While this does not fully converge the exciton energies~\cite{qiu2016screening}, it nonetheless captures the qualitative features of the spectrum sufficient to demonstrate a proof-of-principle of the selection rules derived in \prettyref{sec:selection-rule}~\cite{chan2023giant}.
The lowest four conduction bands and the highest two valence bands are included in the BSE kernel.
In the $GW$-BSE calculation with these parameters, the frequency of the A exciton is \SI{2.10}{eV}, and the frequency of the C exciton is \SI{2.85}{eV} in this calculation.
%A fully converged TD-aGW calculation is left to future works.

To validate our calculation, we start by performing a TD-aGW calculation in which the electric field amplitude is within the linear response regime.
By supplying a narrow gaussian pulse to the system and recording the polarization $P(t)$, the linear absorption spectrum can be calculated using Eq.~\eqref{eq:absorption-2d}.
The result can then be compared with the equilibrium $GW$-BSE absorption spectrum in \prettyref{fig:setup}(b).
We compute the dipole in TD-aGW runs using a finite difference-based approach~\cite{chan2021giant}.
Because the dipole matrix is typically calculated using the momentum-based approach in equilibrium $GW$-BSE software packages~\cite{deslippe2012berkeleygw}, absorption spectra calculated using both methods are presented in  \prettyref{fig:setup}(b).
It can be seen that the equilibrium $GW$-BSE absorption spectrum is consistent with the TD-aGW absorption spectrum.

Following Refs.~\cite{chan2021giant,perfetto2019pump}, We use circularly polarized light for the pump, with a sine wave envelope (\prettyref{fig:setup}(d)).
The amplitude is \SI{2.57e8}{V/m} (or $0.001 / \sqrt{2}$ atomic Rydberg units)
and the pulse duration is \SI{48.4}{fs} (or 1000 atomic Rydberg units).
Coherent polarization in the system persists after the short pump pulse ends (\prettyref{fig:setup}(d)), consistent with previous calculations \cite{chan2021giant}.
The pump frequency controls the species of the predominant coherent excitons.
When the pump frequency is \SI{2.05}{eV}, the A exciton is excited, and only excitons in one valley are excited since the pump is circularly polarized (\prettyref{fig:setup}(d)).
When the pump frequency is \SI{2.80}{eV}, the C exciton is excited, and the lack of valley selection rule means all six valleys at $1/2$K and $1/2$K' (sometimes referred to as the $\Lambda$ or Q valleys), and also the peripheries of both K and K' valleys, see considerable exciton weight (\prettyref{fig:setup}(e)).

\subsection{Results for monolayer \ce{MoS2}}\label{sec:tr-abs-results}

\begin{figure*}
    \centering
    \includegraphics[width=\linewidth]{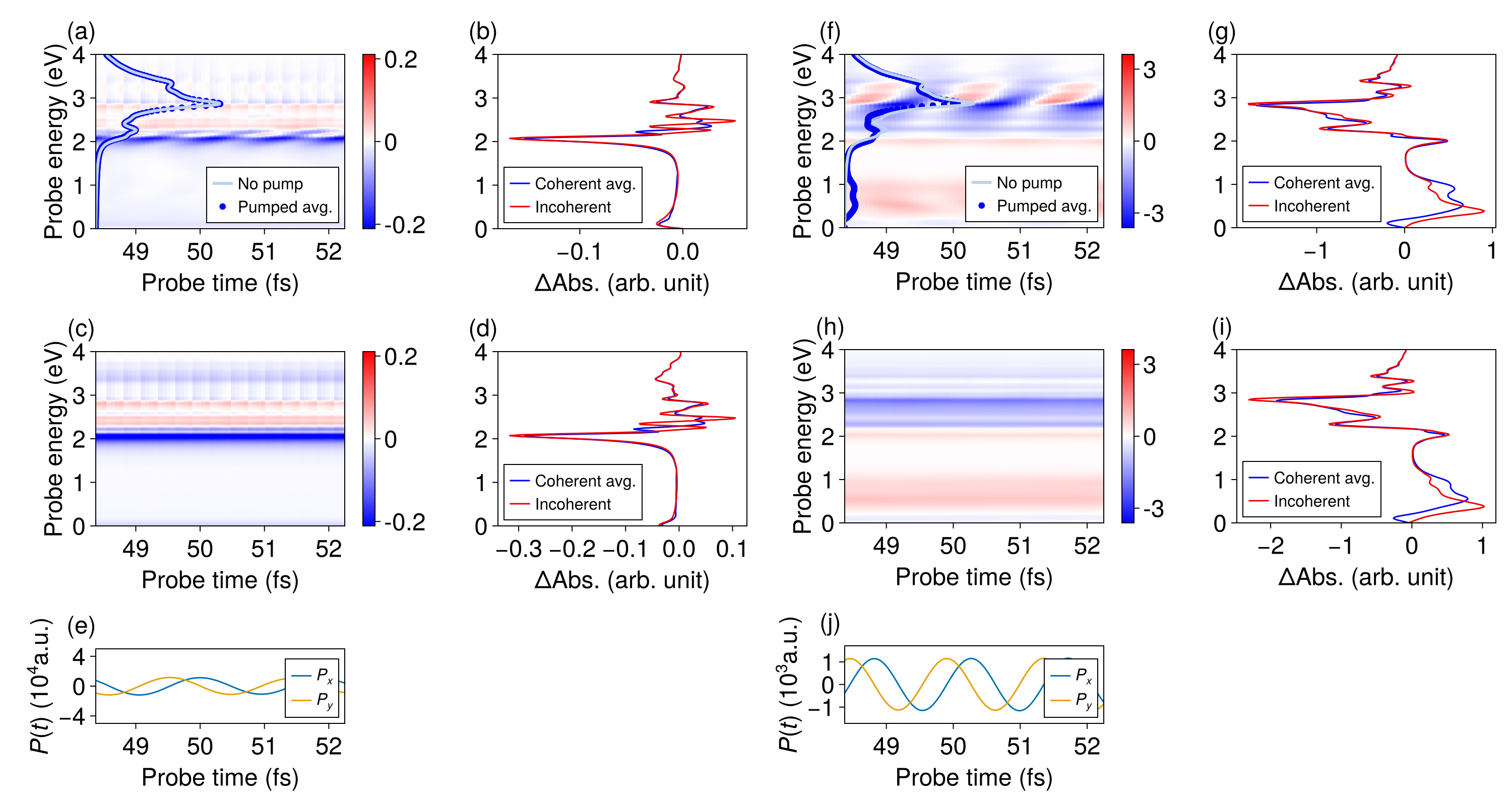}
    \caption{Transient absorption spectra calculated from TD-aGW, compared with the polarization density in the system.
    (a) The difference between the transient absorption spectrum measured along the $y$ axis and the equilibrium linear absorption spectrum calculated using TD-aGW (\prettyref{fig:setup}(b)), with a pump frequency of \SI{2.05}{eV}. The first Brillouin zone distribution of the coherent excitonic oscillation caused by the pump is illustrated in \prettyref{fig:setup}(d).
    %The plots on top of the heatmaps illustrate the non-pumped absorption spectrum and the time-averaged pumped transient absorption spectrum, corresponding to the time-averaged spectrum ($n=0$ component in Eq.~\eqref{eq:absorption-2d-omega}).
    The equilibrium absorption spectrum (solid blue line) and the time-averaged pumped absorption spectrum (blue dots, corresponding to the $n=0$ component in Eq.~\eqref{eq:absorption-2d-omega}) are overlaid along the left edge of the heatmap.
    The colorbar is in arbitrary units.
    (b) Comparison between the the time averaged non-equilibrium renormalization of the transient absorption spectrum (i.e. the difference between the non-pumped and pumped curves in (a)) and the non-equilibrium renormalization calculated with an incoherent initial state obtained by zeroing out the non-diagonal components of the density matrix after the pump.
    (c) Same as (a) but showing transient absorption for a right-circularly polarized probe.
    (d) Same as (b) but with a right-circularly polarized probe.
    (e) The $x$ and $y$ components of the total polarization density in the system, due to the pump.
    The $x$ axis in panels (a-e) corresponds to the probe time, 
    (f-j) Same as (a-e) but with a pump frequency of \SI{2.80}{eV}. The first Brillouin zone distribution of the coherent excitonic oscillation caused by the pump is illustrated in \prettyref{fig:setup}(e).
    }
    \label{fig:tr-abs}
\end{figure*}

Next, we calculate transient absorption by placing a narrow gaussian pulse as the probe pulse with an adjustable delay time from the center of the pump pulse $\tau$,
and evaluate the deviation of the transient absorption spectrum at each delay time from the equilibrium absorption spectrum (\prettyref{sec:method}).
The results are shown in \prettyref{fig:tr-abs}.

In monolayer \ce{MoS2}, the A exciton is restricted to the K valley and is optically bright \cite{qiu2013optical} and therefore is expected to not coherently dress itself at the first-order.
In our simulation, when coherent A excitons are stimulated, in \prettyref{fig:tr-abs}(a), the oscillation frequency of the transient absorption spectrum is twice as much as the frequency of the polarization density.
The time-averaged transient absorption spectrum also demonstrates a less pronounced change, suggesting that excitonic self-driving effects are second-order in this scenario.

\begin{figure}
    \centering
    \includegraphics[width=\linewidth]{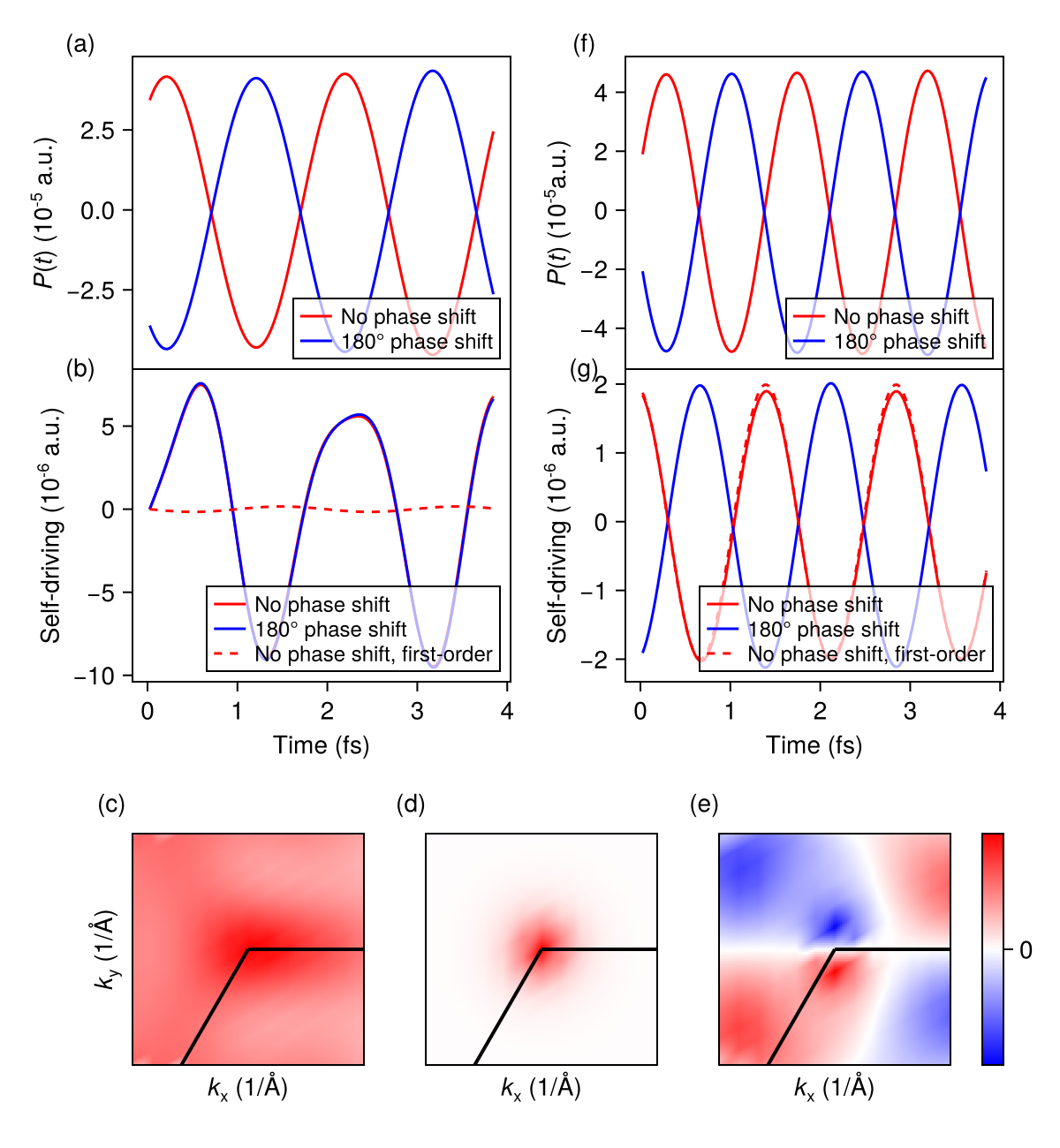}
    \caption{(a) Polarization density obtained by time evolution with an initial state containing coherent A excitons. The coherent exciton initial states corresponding to the red and blue curves have opposite phases.
    (b) The self-driving field defined in Eq.~\eqref{eq:probe-eom} projected to the A exciton wave function.
    The red dashed curve is obtained by replacing $\rho^{\text{pump}}(t)$ with $A^{\text{A}}_{vc\vb{k}} \ee^{-\ii \Omega^{\text{A}} t}$ i.e. by ignoring the higher-order effects in the pump-induced coherent excitonic polarization field, with the appropriate phase factor and amplitude consistent with the initial state of the ``red'' calculation.
    (c) The dipole matrix element between the lowest conduction band and the highest valence band, near the K valley. The gauge is chosen to let the 1s exciton wave function be real everywhere,
    and in this gauge the dipole matrix elements are real.
    (d) The 1s exciton wave function.
    (e) The real part of $K_{c_1 v_1 \vb{k} c_1 c_1 \vb{k}'}$ with $\vb{k}'$ set to the K point.
    The gauge is consistent with that in (c) and (d).
    Note that the value at $\vb{k} = \mathrm{K}$ vanishes,
    and in the K valley, the $GW$-BSE kernel has a $p$-like behavior.
    The imaginary part of the same quantity shows a similar sign structure.
    (f) and (g) are similar to (a) and (b) but the initial states contain coherent C excitons.
    The self-driving field is still projected to the A exciton wave function.
    Note that in (g), the red and blue curves show a phase difference of $\pi$,
    which is absent in (b), and the self-driving field calculated by replacing $\rho^{\text{pump}}(t)$ with $A^{\text{C}}_{vc\vb{k}} \ee^{-\ii \Omega^{\text{C}} t}$, shown as the dashed red curve in (g), is largely consistent with that calculated with a coherent C exciton initial state with the same phase (the red solid curve), which is also not seen in (b).
    These observations suggest that A-to-A coherent excitonic driving is a second-order effect, while C-to-A driving is a first-order effect.}
    \label{fig:no-first-order}
\end{figure}

Because the winding number of the K valley in monolayer \ce{MoS2} is $\pm 1$, $\sum_{\vb{k}} \abs*{A_{v c \vb{k}}^{S}}^2 (K_{cc\vb{k} c v \vb{k}'} - K_{vv\vb{k} c v \vb{k}'})$ has a $p$-wave like sign structure (\prettyref{fig:no-first-order}(e)) in the gauge in which the wave function of the A exciton and the dipole matrix element have a $s$-wave like sign structure (\prettyref{fig:no-first-order}(c, d)).

The absence of first-order excitonic self-driving effects can also be demonstrated by an explicit calculation of the self-driving field.
We perform two TD-aGW calculations with initials states containing coherent spin-allowed 1s A excitons, with a phase difference of $\pi$, producing two $\var{\rho^{\text{pump}}}(t)$ time sequences. Their coherent polarization densities are evaluated and given in \prettyref{fig:no-first-order}(a).
Following the setup in deriving Eq.~\eqref{eq:a-to-a-driving}, 
we evaluate $\comm{\var{\Sigma}^{\text{pump}}}{\var{\rho^{\text{probe}}}} + \comm{\var{\Sigma^{\text{probe}}}}{\var{\rho^{\text{pump}}}}$, projected to the exciton wave function $A^{\text{A}}_{vc\vb{k}}$, in Eq.~\eqref{eq:probe-eom} at different time steps, with $\var{\rho^{\text{pump}}}$ being from numerical TD-aGW simulations, and $\var{\rho^{\text{probe}}}$ set to $A^S_{vc\vb{k}}$.
The results are displayed in \prettyref{fig:no-first-order}(b).
The phase difference in the initial state is expectedly reflected in the polarization density shown in \prettyref{fig:no-first-order}(a).
However, the sign difference is not reflected in the self-driving field felt by A exciton \prettyref{fig:no-first-order}(b), suggesting that the self-driving field is not a first-order response to the coherent A population.
Further, when we replace $\var{\rho}^{\text{pump}}$ by $A^{\text{A}}_{vc\vb{k}} \ee^{-\ii \Omega^{\text{A}} t}$ as is done in deriving Eq.~\eqref{eq:a-to-a-driving} and recalculate the projected self-driving field, the result almost vanishes (\prettyref{fig:no-first-order}(b)).
Thus, self-driving effects felt by the A exciton from coherent A excitons from pumping are not a first-order effect.
This numerically confirms the selection rule in \prettyref{sec:no-a-to-a-renormalization}.

On the other hand, with proliferation of coherent C excitons (\prettyref{fig:tr-abs}(f)), it can be observed that the oscillation frequency of transient absorption is comparable to that of the polarization density, which oscillates at the rate of $\Omega^{\text{C}}$, and not $2 \Omega^{\text{C}}$.
Repeating the procedure calculating the self-driving field in \prettyref{fig:no-first-order}, we note that altering the phase of coherent C excitons in the initial state leads to a sign change in the self-driving effect, 
and after replacing $\var{\rho^{\text{pump}}}$ by $A^{\text{C}}_{vc\vb{k}} \ee^{-\ii \Omega^{\text{C}} t}$, the calculated self-driving field largely reproduces the actual self-driving field from a real TD-aGW time propagation (\prettyref{fig:no-first-order}(f, g)).
Therefore, the leading order of C-to-A excitonic coherent driving is first-order.
This is consistent with the discussion in \prettyref{sec:no-a-to-a-renormalization}, in which we note that the non-trivial band index contraction for multi-band excitons and the C exciton being outside the valleys mean that no obvious cancellation mechanism exists, explaining the presence of first-order C-to-A dressing effects in \prettyref{fig:tr-abs}(f).
We should note that C-to-A coherent dressing may be experimentally challenging to observe, because the lifetime of the C exciton is low. 

We also note that, both in A-to-A coherent dressing and in C-to-A coherent dressing, circularly polarized transient absorption spectra do not oscillate with respect to the delay time, consistent with the selection rule in \prettyref{sec:no-circular-polarization-renormalization}.

Finally, we observe that in both C-to-A and A-to-A driving, if we zero out the non-diagonal components of the density matrix after the pump ends and perform probing on top of this (non-oscillating) incoherent non-equilibrium state, the resulting absorption spectra do not differ considerably from the temporally averaged transient absorption spectrum with coherent initial states (\prettyref{fig:no-first-order}).
This indicates that at least in the case of monolayer \ce{MoS2}, coherent excitonic dressing is most uncontroversially identified by the time dependence of the transient absorption spectrum, and not by the patterns of excitonic peak height changes and frequency shifts.

\subsection{Strained monolayer \ce{MoS2}}

\begin{figure*}
    \centering
    \includegraphics[width=\linewidth]{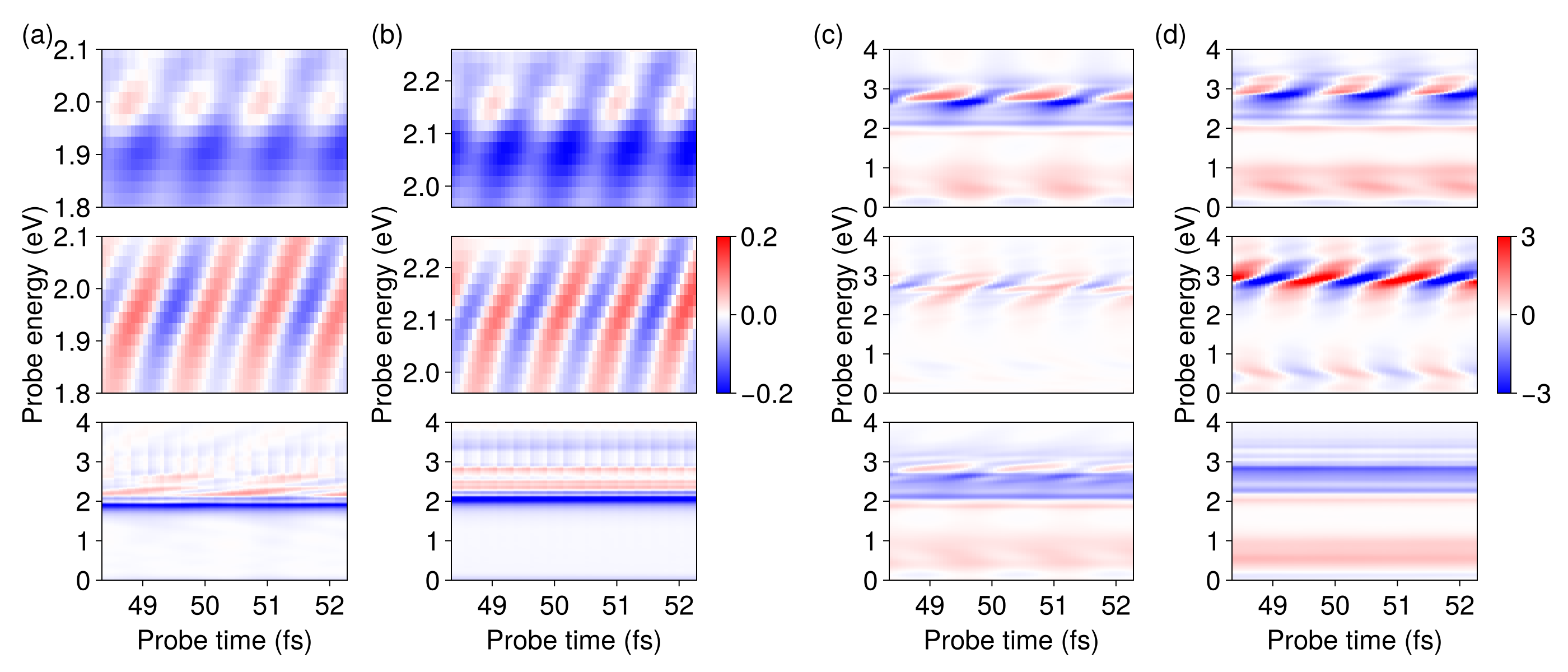}
    \caption{Comparison between transient absorption with and without structural strain. The three rows visualize non-equilibrium renormalization of the transient absorption spectrum along the $y$ axis (top row), non-equilibrium renormalization of $\Im (\chi_{xy} + \chi_{yx})$ (center), and non-equilibrium renormalization of the right circularly polarized transient absorption spectrum (bottom).
    (a) \ce{MoS2}, \SI{3}{\percent} strained along the $y$ direction, with pump resonant with A peak.
    (b) Unstrained \ce{MoS2} with pump resonant with A peak.
    (c) \ce{MoS2}, \SI{3}{\percent} strained along the $y$ direction, with pump resonant with C peak.
    (d) Unstrained \ce{MoS2} with pump resonant with C peak.
    }
    \label{fig:strain}
\end{figure*}

Both selection rules in \prettyref{sec:selection-rule} rely on the $C_3$ symmetry of monolayer \ce{MoS2}.
In  \prettyref{fig:strain}, we apply a \SI{3}{\percent} uniform uniaxial strain to the system along the $y$ direction.
This removes the $C_3$ symmetry and therefore removes the selection rules.
We observe that after applying the strain, strict frequency doubling with A driving breaks down (\prettyref{fig:strain}(a)), and circularly polarized transient absorption spectra also have oscillating features (\prettyref{fig:strain}(a, c)).

\section{Conclusion}\label{sec:conclusion}

In this paper, we have formulated a comprehensive analysis of TD-aGW and reveal that TD-aGW treats non-equilibrium coherent dressing effects in pump-probe spectroscopy as a classical nonlinear field with the nonlinearity coming from the $GW$-BSE kernel.
We have demonstrated that tr-ARPES and transient absorption receive different non-equilibrium coherent renormalization. 
Further, we have shown that coherent exciton dressing of transient absorption follows selection rules originating from macroscopic symmetries and symmetry-protected microscopic band structure topology, which are not seen in coherent exciton dressing effects in ARPES. 
This means that, in monolayer systems with $C_n$ symmetry, exciton self-driving does not produce first-order renormalization of circularly polarized transient absorption, while in high-symmetry, topologically nontrivial valleys, optically active exciton modes do not coherently dress one another at first order. These observations are numerically demonstrated in the excitonic material monolayer \ce{MoS2}. 
We further show that breaking the $C_n$ symmetry lifts the selection rules.

\section{Acknowledgements}

This work was primarily supported by the National Science Foundation (NSF) Condensed Matter and Materials Theory (CMMT) program under Career Grant Number DMR-2337987. Development of the BerkeleyGW code was supported by Center for Computational Study of Excited-State Phenomena in Energy Materials (C2SEPEM) at the Lawrence Berkeley National Laboratory, funded by the U.S. Department of Energy, Office of Science, Basic Energy Sciences, Materials Sciences and Engineering Division, under Contract DE-AC02-05CH11231. The calculations used resources of the National Energy Research Scientific Computing (NERSC), a DOE Office of Science User Facility operated under contract no. DE-AC02-05CH11231 under award numbers BES-ERCAP-0031507 and BES-ERCAP-0027380. An award of computer time was provided by the U.S. Department of Energy’s (DOE) Innovative and Novel Computational Impact on Theory and Experiment (INCITE) Program. This research used supporting resources at the Argonne and the Oak Ridge Leadership Computing Facilities. The Argonne Leadership Computing Facility at Argonne National Laboratory is supported by the Office of Science of the U.S. DOE under Contract No. DEAC02-06CH11357 . The Oak Ridge Leadership Computing Facility at the Oak Ridge National Laboratory is supported by the Office of Science of the U.S. DOE under Contract No. DE-AC05-00OR22725.

\appendix

\section{Theoretical details of transient absorption}\label{appx:theory-absorption-detail}

In equilibrium absorption spectroscopy, the absorption spectrum is proportional to the imaginary part of the dielectric function $\epsilon_2(\omega)$.
%However, in the theory of non-equilibrium spectroscopy, as the linear response function has a dependence on two times, a well-defined $\omega$ is absent for the most generic situation due to the time translational symmetry being broken in the non-equilibrium state.
In non-equilibrium systems, however, the linear response function depends on two independent times because time-translation symmetry is broken, and the optical response can no longer be characterized by a single frequency.
In realistic transient absorption measurements, though, the probe pulse is typically a wave packet with a well-defined frequency profile, centered at $\omega$, with a pump-probe delay time $\tau$ that is varied, while the frequency profile and temporal envelope of the probe fixed remain fixed.
The transient absorption spectrum can therefore be defined as the energy loss of the transmitted probe, which depends on both the probe frequency  $\omega$ and the time delay $\tau$.

When the length scale of the sample in the direction of light propagation is small enough so that delay effects due to finite light speed are ignorable, by Maxwell's equations, the relation between the electric field of the probe pulse in free space $\vb{e}(t)$ and the transmitted electric field $\vb{e}'(t)$ is given by \cite{perfetto2015some}
\begin{equation}
    \vb{e}'(t) = \vb{e}(t) + \frac{2\pi}{Sc} \dv{t} \vb{d}(t),
    \label{eq:e-and-ep}
\end{equation}
where $S$ is the projection of the surface of the sample in the wave propagation direction,
and $\vb{d}(t)$ is the averaged dipole, i.e. polarization density, induced by the probe field in the material.
In a system with a non-equilibrium back ground, $\vb{d}$ should be understood as the change brought by the probe field on top of possible existing dipole oscillation.
$\vb{d}(t)$ can be explicitly calculated from the electric field \emph{in} the material
(while both $\vb{e}(t)$ and $\vb{e}'(t)$ are outsides the system).
The absorption spectrum, defined by energy loss in the material sample, is calculated by definition as
\begin{equation}
    \begin{aligned}
        \text{absp. at $\omega$} &= \frac{
        \int \dd{\omega} (\abs{\vb{e}(\omega)}^2 - \abs{\vb{e}'(\omega)}^2)
        }{
            \int \dd{\omega} \abs{\vb{e}(\omega)}^2
        }  \\
        &\simeq \frac{\Delta \omega (\abs{\vb{e}(\omega)}^2 - \abs{\vb{e}'(\omega)}^2)}{\Delta \omega \abs{\vb{e}(\omega)}^2} \\
        &= - \frac{4\pi}{Sc} \Im \frac{\omega \vu{e}^* \cdot \vb{d}(\omega)}{e(\omega)}
        - \frac{4\pi^2}{S^2 c^2} \omega^2 \frac{\abs{\vb{d}(\omega)}^2}{\abs{\vb{e}(\omega)}^2},
    \end{aligned}
\end{equation}
where $\vu{e}$ is the polarization direction of the probe and thus $\vb{e}(\omega) = \vu{e} e(\omega)$.
Here we have made use of the assumption that the frequency profile of the pulse is narrow.
Assuming that the $\vb{d}^2$ term can be discarded,
which is not generally correct for thick materials \cite{perfetto2015some} but is acceptable for two-dimensional systems \cite{perfetto2015nonequilibrium},
the absorption spectrum is given by 
\begin{equation}
    \text{absp. at $\omega, \tau$} \sim \Im \frac{\vu{e}^* \cdot \vb{d}(\omega)}{e(\omega)} ,
    \label{eq:absorption-2d-pre}
\end{equation}
where the dependence on the delay time $\tau$ is implicitly introduced in
$\vb{d}(\omega)$ and $\vb{e}(\omega)$,
as it controls the temporal position of the wave packet in $\vb{e}(t)$
and hence $\vb{d}(t)$.

A caveat in \eqref{eq:absorption-2d-pre} is that $\vb{e}(t)$ is the \emph{incident} electric field,
and not the total electric field within the system \cite{perfetto2015some}.
The linear response of $\vb{d}(t)$ to $\vb{e}(t)$ therefore is the linear response of $\vb{d}(t)$ to the total electric field times the ratio between the total electric field and the external field.
For thin samples, however, the second part does not typically introduce additional pole structures,
as the polarization directions of optical bound states at the surface of the system typically are perpendicular to the surface, not parallel to it.
Thus, the pole structure contribution to the absorption spectrum can be evaluated as
\begin{equation}
    \text{absp. at $\omega, \tau$} \sim \Im \frac{\vu{e}^* \cdot \vb{d}(\omega)}{{E}(\omega)} ,
\end{equation}
which is \eqref{eq:absorption-2d} in the main text.

Unlike the equilibrium dielectric function, \eqref{eq:absorption-2d} depends on $\vb{e}(t)$, although not on its amplitude.
The most important consequence is \eqref{eq:absorption-2d} depends on the pump-probe delay time $\tau$.
In particular, when the system undergoes periodic driving,
we expect \eqref{eq:absorption-2d} to oscillate periodically as $\tau$ varies,
with the same period of the driving field.
This is because the response function of a Floquet system driven by a field with frequency $\omega_0$ has the general form of \cite{rudner2020floquet}
\begin{equation}
    \chi(t, t') = \sum_{m} \int_{-\infty}^\infty \frac{\dd{\omega}}{2\pi} \ee^{-\ii \omega (t - t')} \ee^{-\ii m \omega_0 t'} \chi^{(m)}(\omega),
\end{equation}
which leads to \eqref{eq:absorption-2d-omega} in the main text.

\section{Time dependent $GW$}\label{appx:td-gw}

Similar to many-body perturbation theory in equilibrium systems, in non-equilibrium systems, the equation of motion of the single-particle Green's function can still be evaluated using the Dyson equation with self-energy correction,
which, following the Langreth rules, is equivalent to a set of self-consistent equations for the retarded, advanced, lesser, and greater double-time Green's functions, known as the Kadanoff-Baym equations (KBE)
\cite{kremp2005quantum,kadanoff2018quantum},
from which TD-aGW can be derived by further imposing the static COHSEX approximation \cite{attaccalite2011real,perfetto2015nonequilibrium,sangalli2021excitons,chan2021giant,chan2023giant}.
When the adiabatic approximation is not made \cite{perfetto2022real}, the tree-like diagrammatic structure of TD-aGW remains identical, but the exciton diagrams are now also double-timed.
Periodic self-driving by a double-timed self-energy is currently theoretically underexplored.
We leave the issue to future studies.

We also note that all excitons mentioned in this work are coherent excitons, and a faithful description of ``real'' excitons needs diagrams beyond $GW$, adiabatic or not \cite{perfetto2016first},
omitted in this work since we are mostly interested in coherent excitonic effects.

The electric field $\vb{E}$ in TD-aGW simulations in this work is the total macroscopic electric field in the system,
which already includes macroscopic radiation from the electrons.
This means the long-range exchange part of the $GW$-BSE kernel should be zeroed out \cite{hou2025data}.
We have made the usual approximation and assume that the characteristic wave length of $\vb{E}$ is much larger than the characteristic length scale of the material (typically the lattice constants), and thus the crystal momenta of electrons are approximately conserved even in non-equilibrium processes,
causing both $\rho$ and $\var{\Sigma}^{GW}$ to have only one momentum index.

\section{$\vb{k} \cdot \vb{p}$ analysis of the generic low-energy Hamiltonian}\label{appx:kphamiltonian}

In this Appendix, we prove that in a valley with $C_n$ symmetry, in the region sufficiently close to the center of the valley, Eq.~\eqref{eq:two-band-ham} is the most generic form of the low-energy Hamiltonian.
We also derive the winding $w$ and show why $w=\pm1$ in the K and K' valleys in monolayer \ce{MoS2}.

Let $\mathbf{K}_\tau$ ($\tau=\pm 1$) denote the center of the valley under consideration, and let $\bfk$ denote the momentum measured relative to the valley center.
We project the quasiparticle Hamiltonian onto a valence- and conduction-band pair near the band edge $\{\ket{u^\tau_{c\bfk=0}},\ket{u^\tau_{v\bfk=0}}\}$.
In this fixed basis, the most general two-band Hamiltonian can be written as
\begin{equation}
    H_{\text{eff}}(\bfk) = d_0(\bfk) \sigma_0 + \mathbf{d}(\bfk)\cdot\boldsymbol{\sigma}.
\end{equation}
For later convenience, we define $f(\bfk) = (d_x(\bfk) - \ii d_y(\bfk))/2$ and $\sigma_{\pm}=\sigma_x\pm\ii\sigma_y$.
Because $\{\ket{u^\tau_{c\vb{k}=0}},\ket{u^\tau_{v\vb{k}=0}}\}$ are chosen to diagonalize the Hamiltonian at the valley center, so $f(0)=0$ and $d_z(0) = \Delta$ where $2\Delta$ is the direct band gap.
Then, the effective Hamiltonian can be written as
\begin{equation}\label{eq:chiral_ham}
    H_{\text{eff}}(\bfk) = d_0(\vb{k}) \sigma_0 + d_z(\bfk)\sigma_z + f(\bfk)\sigma_{+} + f^*(\bfk)\sigma_{-}.
\end{equation}
The angular dependence of $f(\bfk)$ is fixed by the rotational representations of $\{\ket{u^\tau_{c\vb{k}=0}},\ket{u^\tau_{v\vb{k}=0}}\}$.
Let
\begin{equation}
    C_n \ket{u^\tau_{i0}} = \ee^{2\pi \ii \tau m_i/n}\ket{u^\tau_{i0}},\quad i=c,v,
\end{equation}
where $m_i\in\mathbb{Z}_n$ is determined by the wavefunction symmetry of the band-edge states (e.g., for monolayer \ce{MoS2}~\cite{xiao2012coupled}, the conduction band is $\ket{d_{z^2}}$ and the valence band is $\ket{d_{x^2-y^2}}+\ii\tau \ket{d_{xy}}$ so we have $m_c=0$ and $m_v = 1$.)
Therefore, the representation matrix $U_n$ of the $C_n$ rotation under the basis $\{\ket{u^\tau_{c\vb{k}=0}},\ket{u^\tau_{v\vb{k}=0}}\}$ can be written as
\begin{equation}
    U_n = \pmqty{\dmat{\ee^{2\pi \ii \tau m_c /n}, \ee^{2\pi\ii\tau m_v /n}}}.
\end{equation}
Then, we have $U_n \sigma_{\pm} U_n^\dagger = \ee^{\pm 2\pi\ii \tau(m_c-m_v)/n}\sigma_{\pm}$ and $U_n\sigma_z U_n^\dagger = \sigma_z$.
The effective Hamiltonian transforms covariantly under $C_n$ rotation, namely
\begin{equation}
    H_{\text{eff}}(C_n \bfk) = U_n H_{\text{eff}}(\bfk) U_n^\dagger.
\end{equation}
Plugging Eq.~\eqref{eq:chiral_ham} into the right hand side of the equation, we get
\begin{equation}
\begin{aligned}
    H_{\text{eff}}(C_n\bfk) &= d_0(\vb{k}) \sigma_0 + d_z(\bfk)\sigma_z  \\
    &+ \ee^{2\pi\ii \tau(m_c-m_v)/n}f(\bfk)\sigma_+ \\
    &+ \ee^{- 2\pi\ii \tau(m_c-m_v)/n}f^*(\bfk)\sigma_{-}.
\end{aligned}
\end{equation}
Comparing the coefficients, we get the following constraints as a result of the $C_n$ symmetry of the $\mathbf{K}_\tau$ valley:
\begin{align}
    \label{eq:C3_constraint_on_f}f(C_n \bfk) &= \ee^{2\pi\ii\tau(m_c- m_v)/n} f(\bfk), \\
    \label{eq:C3_constraint_on_dz}d_z(C_n \bfk) &= d_z(\bfk), \quad d_0(C_n \bfk) = d_0(\bfk).
\end{align}
The constraint can be made explicit by introducing $k_{\pm} = k_x\pm \ii k_y$, which transform as $k_{\pm}\to \ee^{\pm 2\pi\ii /n} k_\pm$ under $C_n$.
In general, the $\vb{k} \cdot \vb{p}$ Hamiltonian coefficients can be expanded into polynomials
\begin{equation}
    f(\bfk) = \sum_{\mu\nu}c_{\mu\nu}k^\mu_{+} k^\nu_{-}.
\end{equation}
Here $\mu, \nu \geq 0$ to avoid singularity.
Under $C_n$ rotation, we have
\begin{equation}
    f(C_n \bfk) = \sum_{\mu\nu}\ee^{2\pi\ii(\mu-\nu)/n}c_{\mu\nu}k^\mu_{+} k^\nu_{-}.
\end{equation}
In the symmetric constraint Eq.~\eqref{eq:C3_constraint_on_f}, a coefficient $c_{\mu\nu}$ can be nonzero if and only if
$\mu-\nu = \tau(m_c-m_v) + ni$ where $i \in \mathbb{Z}$.
The constant 
\begin{equation}
    \tau(m_c-m_v) \eqcolon w
\end{equation}
is an integer decided by the angular momentum of Block states at the center of the valley, and is the winding number in Eq.~\eqref{eq:two-band-ham} modulo $n$.
Since $k_\pm=|\bfk| \ee^{\pm\ii \phi_\bfk}$ where $\phi_\bfk=\arg\bfk$, the off diagonal term in the effective Hamiltonian now reads
\begin{equation}
    \begin{aligned}
        f(\bfk) &= \sum_{\mu \nu} c_{\mu \nu} \abs{\vb{k}}^{\mu+\nu} \ee^{\ii (\mu - \nu) \phi_{\vb{k}}} \\
        &= \sum_{\nu i} c_{\mu = \nu + w + ni, \nu} |\bfk|^{2\nu+w+ni} \ee^{\ii(w+ni)\phi_\bfk}.
    \end{aligned}
    \label{eq:expansion-of-f}
\end{equation}
%Now, assuming that $\abs{\vb{k}}$ is sufficiently small, the $i \neq 0$ terms are always of a higher order than the $i=0$ terms, and thus can be ignored.
For optically active valleys with $C_n$ symmetry ($n\geq3$), $w=\pm1$ mod $n$, so the $i=0$ term is the leading symmetry-allowed contribution, assuming its coefficient is nonzero. The higher-order terms can therefore be neglected sufficiently close to the valley center.
Hence 
\begin{equation}
    f(\vb{k}) = \ee^{\ii w \phi_{\vb{k}}} \alpha(|\bfk|), \quad \alpha(|\bfk|)=\sum_{\mu\nu}c_{\mu\nu}|\bfk|^{\mu+\nu},
\end{equation}
and we get exactly the same off-diagonal term in Eq.~\eqref{eq:two-band-ham}.
A similar analysis shows that in a neighborhood small enough of the valley center, $d_0(\vb{k}) = d_0(\abs{\vb{k}})$, and $d_z(\vb{k}) = d_z(\abs{\vb{k}})$.
Thus, we proved the effective Hamiltonian shown in Eq.~\eqref{eq:two-band-ham}.
Intuitively, this means the discrete rotational symmetry is enough to induce an emergent continuous rotational symmetry to leading order near the valley.

When $n=2$, it is possible to have multiple terms in Eq.~\eqref{eq:expansion-of-f} with different $(\nu, i)$ but the same order with respect to $\abs{\vb{k}}$.
An example is $f(\vb{k}) = \abs{\vb{k}} (a \ee^{\ii \phi_{\vb{k}}} + b \ee^{-\ii \phi_{\vb{k}}})$, where $w = -1$ and $(\nu, i) = (1, 0), (0, 1)$.
The structure of the single-electron Hamiltonian is therefore much more complex in this case and is not analyzed in this work.

\section{Symmetry of the $GW$-BSE kernel}\label{appx:symmetry-gw-bse}

The symmetry representations carried by wave functions at the center of the valley and the emergent continuous rotational symmetry in Eq.~\eqref{eq:two-band-ham} has consequences for both the symmetry of the $GW$-BSE kernel in momentum space \cite{bajaj2025symmetries} and also its sign structure.

The $GW$-BSE kernel can be decomposed into a direct part and an exchange part, 
involving the screened Coulomb interaction and the bare Coulomb interaction, respectively \cite{hou2025data,deslippe2012berkeleygw}.
Each Bloch wave function $\psi_{m\vb{k}}(\vb{r})$  appearing in the basis contains both the Bloch phase factor $\ee^{\ii \vb{k} \cdot \vb{r}}$ and the periodic part $u_{n\vb{k}}(\vb{r})$.
In equilibrium $GW$-BSE \cite{rohlfing2000electron,deslippe2012berkeleygw}
\begin{equation}
    (E_{c\vb{k}} - E_{v\vb{k}}) A^S_{vc\vb{k}} + \sum_{c' v' \vb{k}'} K_{cv\vb{k} c'v'\vb{k}'} A^S_{v'c'\vb{k}'} = \Omega_S A^S_{vc\vb{k}},
\end{equation}
only matrix elements where $m$, $m'$ are in the conduction band manifold and $n$, $n'$ are in the valence band manifold are involved.
%Assuming exciton formation in the valley, these matrix elements should be non-zero in most places in the valley.
%In this scenario, because the periodic part contains $\vb{G} \neq 0$ momentum components only, when $\vb{k}$ is small compared with the size of the first Brillouin zone, we can approximate $\psi_{m\vb{k}}(\vb{r})$ by $u_{m\vb{k}=0} \ee^{\ii \vb{k} \cdot \vb{r}}$.
For momenta sufficiently close to the valley center, we approximate the periodic Bloch functions by their values at the valley center,  such that $\psi_{m\mathbf{k}}(\mathbf r)\approx u_{m,\mathbf{k}=0}(\mathbf r)\ee^{\ii\mathbf{k}\cdot\mathbf r}$.
In this way, when $\vb{k}$ is small compared with the size of the first Brillouin zone, the direct term reads 
\begin{widetext}
    \begin{equation}
    \begin{aligned}
        K^{\text{d}}_{mn\vb{k} m'n'\vb{k}'} &= \int \dd{\vb{r}} \int \dd{\vb{r}'} \psi^*_{m\vb{k}}(\vb{r}) \psi_{m'\vb{k}'}(\vb{r}) W(\vb{r}, \vb{r}') \psi_{n\vb{k}}(\vb{r}') \psi^*_{n'\vb{k}'}(\vb{r}')   \\
        &\approx  \int \dd{\vb{r}} \ee^{\ii \vb{r}  \cdot (\vb{k}' - \vb{k})} \int \dd{\vb{r}'}  \ee^{- \ii \vb{r}' \cdot (\vb{k}' - \vb{k})} \underbrace{u^*_{m\vb{k}=0}(\vb{r}) u_{m'\vb{k}'=0}(\vb{r}) W(\vb{r}, \vb{r}') u_{n\vb{k}=0}(\vb{r}') u^*_{n'\vb{k}'=0}(\vb{r}')}_{\eqcolon \tilde{W}(\vb{r}, \vb{r}')},
    \end{aligned}
    \label{eq:k-d}
    \end{equation}
\end{widetext}
where $W(\vb{r}, \vb{r}')$ is the screened Coulomb interaction in real space, and the exchange part
\begin{widetext}
    \begin{equation}
    \begin{aligned}
        K^{\text{x}}_{mn\vb{k} m'n'\vb{k}'} &= \int \dd{\vb{r}} \int \dd{\vb{r}'} \psi_{m\vb{k}}^*(\vb{r}) \psi_{n\vb{k}}(\vb{r}) v(\vb{r} - \vb{r}') \psi_{m'\vb{k}'}(\vb{r}') \psi^*_{n'\vb{k}'}(\vb{r}') \\ 
        &= \int \dd{\vb{r}} \int \dd{\vb{r}'} u_{m\vb{k}=0}^*(\vb{r}) u_{n\vb{k}=0}(\vb{r}) v(\vb{r} - \vb{r}') u_{m'\vb{k}'=0}(\vb{r}') u^*_{n'\vb{k}'=0}(\vb{r}').
    \end{aligned}
    \label{eq:k-x}
    \end{equation}
\end{widetext}
We note that the integrands in Eqs.~\eqref{eq:k-d} and \eqref{eq:k-x}, ignoring the Fourier factors, carry representations of the $C_n$ group.
For two-band excitons, the only interacting term in $GW$-BSE is $K_{cv\vb{k}cv\vb{k}'}$.
This means the phase factors of the representations of the $C_n$ group carried by the four $u_{n\vb{k}}$ functions cancel.
Thus, the integrand $\tilde{W}(\vb{r}, \vb{r}')$ in Eq.~\eqref{eq:k-d} is invariant under $C_n$ rotation. 
Because in Eq.~\eqref{eq:k-d}, $\vb{k}$ and $\vb{k}'$ dependences primarily come from the Fourier factor, 
this in turn means that $K^{\text{d}}_{cv\vb{k}cv\vb{k}'}$ is invariant under $C_n$ rotation of both $\vb{k}$ and $\vb{k}'$ at the same time.
Similarly, the integrand of Eq.~\eqref{eq:k-x} is invariant under $C_n$ rotation, and $K^{\text{x}}_{cv\vb{k}cv\vb{k}'}$ in fact has weak dependence on $\vb{k}$ and $\vb{k}'$.
Therefore, when $\vb{k}$ is sufficiently small, $K_{cv\vb{k}cv\vb{k}'}$ is invariant under $C_n$ rotation.
More generally, the BSE Hamiltonian under the Tamm-Dancoff approximation \cite{rohlfing2000electron} is invariant under symmetric operations that apply to both the electron and the hole \cite{bajaj2025symmetries}.

This means that in a valley with $C_n$ point group, in the two-band scenario, the $GW$-BSE kernel preserves the angular momentum modulo $n$ of the exciton wave function.
Thus, the wave function of any two-band exciton mode in the valley should at least have a well-defined angular momentum quantum number modulo $n$, provided that the exciton is localized enough in the valley (which is the case for monolayer \ce{MoS2}; \prettyref{fig:no-first-order}(d, e)).
In monolayer \ce{MoS2}, the K-valley exciton wave functions approximately have separated radial and angular dependence and well-defined angular momentum quantum numbers, with both $s$-like and $f$-like states \cite{qiu2016screening}.
In monolayer hexagonal \ce{BN}, the amplitude of the exciton wave function has angle dependence, but an angular momentum quantum number modulo three can still be recognized \cite{zhang2022intervalley}.
Regardless of the angle dependence of the amplitude of the exciton wave function, by applying the $C_n$ rotation, it follows that if the two functions $f$ and $g$ have different angular momenta modulo $n$, their overlap integral vanishes.

We emphasize that the argument above is about the exciton envelope angular momentum, describing the angular dependence of $A^S_{vc\mathbf{k}}$ with respect to the internal momentum $\vb{k}$. It is distinct from the angular momentum carried by the band-edge Bloch states, while the total exciton angular momentum includes both contributions. Although the exciton wave functions and the GW-BSE kernel are gauge-dependent, angular momentum conservation is gauge-independent.
%We emphasize that this angular momentum is the exciton envelope angular momentum, which is shown by the exciton wave function $A^S_{vc\vb{k}}$ having a phase factor $\ee^{\ii m \phi_{\vb{k}}}$ with respect to $\vb{k}$ (which is a good quantum number of a single quasiparticle but is not a good quantum number and corresponds to the internal structure of an exciton), and is different from the single-quasiparticle (i.e. Bloch) angular momentum carried by $\ket{c\vb{k}=0}$ and $\ket{v\vb{k}=0}$, which is manifested by the phase structure of the single-quasiparticle wave function with respect to $\vb{r}$ (and not the exciton internal momentum $\vb{k}$).
%It is also different from the total orbital angular momentum of the exciton, which is the sum of the angular momenta from the exciton wave function $A^S_{vc\vb{k}}$ and from the single-quasiparticle Bloch states.
Also, although the exciton wave functions and the $GW$-BSE kernel (when viewed as matrices acting on the excitonic Hilbert space) are gauge-dependent, the rule of angular momentum conservation is gauge independent.

Now we turn to the $K_{cccv}$ and $K_{vvcv}$ matrix elements involved in non-linearity \cite{hou2025data}.
We note that under the approximations in Eqs.~\eqref{eq:k-d} and \eqref{eq:k-x},
these matrix elements vanish at $\vb{k} = \vb{k}' = 0$, where angular momentum conservation strictly holds, and the angular momentum difference between $c$ and $v$ in Eq.~\eqref{eq:band-angular-momentum-dipole-rule} means the transitions in $K_{ccvc}$ and $K_{vvvc}$ are forbidden.
Therefore, the leading-order approximation of $K_{cccv}$ and $K_{vvcv}$ matrix elements has to involve the smaller components of $u_{c/v\vb{k}}$ in Eq.~\eqref{eq:eigenstates-mos2}.
The integral (and the absence of the Fourier factor) in the exchange term enforces angular momentum conservation modulo $n$, and the non-zero terms in $K_{cccv}$ read
\begin{widetext}
    \begin{equation}
    \begin{aligned}
        K^{\text{x}}_{c c \vb{k} c v \vb{k}'} &= \int \dd{\vb{r}} \int \dd{\vb{r}'} \abs{\psi_{c\vb{k}} (\vb{r})}^2 v(\vb{r} - \vb{r}') \psi_{c\vb{k}'}(\vb{r}') \psi^*_{v\vb{k}'}(\vb{r}') \\
        &\approx - \int \dd{\vb{r}} \int \dd{\vb{r}'} \abs{u_{c\vb{k}=0}(\vb{r})}^2 v(\vb{r} - \vb{r}') \abs{u_{c\vb{k}=0}(\vb{r}')}^2  \frac{\theta_{\vb{k}'}}{2} \ee^{-\ii w \phi_{\vb{k}'}} \\
        &+  \int \dd{\vb{r}} \int \dd{\vb{r}'} \abs{u_{c\vb{k}=0}(\vb{r})}^2 v(\vb{r} - \vb{r}') \abs{u_{v\vb{k}=0}(\vb{r}')}^2  \frac{\theta_{\vb{k}'}}{2} \ee^{- \ii w \phi_{\vb{k}'}} \\
        &+ \int \dd{\vb{r}} \int \dd{\vb{r}'} u_{c\vb{k}=0}^*(\vb{r}) u_{v\vb{k}=0}(\vb{r}) v(\vb{r} - \vb{r}') u_{c\vb{k}'=0}(\vb{r}') u^*_{v\vb{k}'=0}(\vb{r}') \frac{\theta_{\vb{k}}}{2} \ee^{- \ii w \phi_{\vb{k}}} \\
        &+ \int \dd{\vb{r}} \int \dd{\vb{r}'} u_{v\vb{k}=0}^*(\vb{r}) u_{c\vb{k}=0}(\vb{r}) v(\vb{r} - \vb{r}') u_{c\vb{k}'=0}(\vb{r}') u^*_{v\vb{k}'=0}(\vb{r}') \frac{\theta_{\vb{k}}}{2} \ee^{\ii w \phi_{\vb{k}}}.
    \end{aligned}
    \label{eq:k-x-cccv}
    \end{equation}
\end{widetext}
Each term in $K_{cccv}^{\text{x}}$ contains a $\ee^{-\ii w \phi_{\vb{k}}}$ factor or a $\ee^{- \ii w \phi_{\vb{k}'}}$ factor.
Therefore, the exchange part of the $GW$-BSE kernel permits only $w \to 0$ and $0 \to -w$ angular momentum (with respect to the momentum) transitions.

Proving a similar rule for $K^{\text{d}}$ is more complicated because the Fourier factors depend on $\vb{q} = \vb{k} - \vb{k}'$.
The first-order contributions to $K^{\text{d}}_{cc\vb{k} cv\vb{k}'}$ with respect to $\vb{k}$ and $\vb{k}'$ now read 
\begin{widetext}
    \begin{equation}
    \begin{aligned}
        K^{\text{d}}_{cc\vb{k} cv\vb{k}'} &\approx \int \dd{\vb{r}} \ee^{\ii \vb{r} \cdot (\vb{k}' - \vb{k})} \int \dd{\vb{r}'} \ee^{\ii \vb{r}' \cdot (\vb{k} - \vb{k}')} u_{c\vb{k}=0}^*(\vb{r}) u_{c\vb{k}'=0}(\vb{r}) W(\vb{r}, \vb{r}') u_{c\vb{k}=0}(\vb{r}') u^*_{v\vb{k}'=0}(\vb{r}')  \\
        &- \int \dd{\vb{r}} \int \dd{\vb{r}'} u_{c\vb{k}=0}^*(\vb{r}) u_{c\vb{k}'=0}(\vb{r}) W(\vb{r}, \vb{r}') u_{c\vb{k}=0}(\vb{r}') u^*_{c\vb{k}'=0}(\vb{r}') \frac{\theta_{\vb{k}'}}{2} \ee^{-\ii w \phi_{\vb{k}'}} \\ 
        &+ \int \dd{\vb{r}} \int \dd{\vb{r}'} u_{c\vb{k}=0}^*(\vb{r}) u_{c\vb{k}'=0}(\vb{r}) W(\vb{r}, \vb{r}') u_{v\vb{k}=0}(\vb{r}') u^*_{v\vb{k}'=0}(\vb{r}') \frac{\theta_{\vb{k}}}{2} \ee^{-\ii w \phi_{\vb{k}}} \\
        &+ \int \dd{\vb{r}} \int \dd{\vb{r}'} u_{c\vb{k}=0}^*(\vb{r}) u_{v\vb{k}'=0}(\vb{r}) W(\vb{r}, \vb{r}') u_{c\vb{k}=0}(\vb{r}') u^*_{v\vb{k}'=0}(\vb{r}') \frac{\theta_{\vb{k}'}}{2} \ee^{-\ii w \phi_{\vb{k}'}}.
    \end{aligned}
    \end{equation}
\end{widetext}
Here the last three terms come from the minor components in Eq.~\eqref{eq:eigenstates-mos2}. These terms do not vanish in the $\vb{q} \to 0$ limit, and therefore the $\vb{q}$ dependence of the Fourier factor is ignored under the first-order approximation.
These terms, like terms in $K^{\text{x}}_{cc\vb{k} cv\vb{k}'}$ in Eq.~\eqref{eq:k-x-cccv}, only contain $w \to 0$ and $0 \to -w$ angular momentum transition channels with respect to the momentum index. 
The first term vanishes when $\vb{q}=0$ due to angular momentum conservation, but may have a finite value away from $\vb{q}=0$, and thus the Fourier factor should be kept.
This term depends on $\vb{q}$ only and its symmetric behavior is therefore comparable to that of Eq.~\eqref{eq:k-d}. However, as is stated in the main text, in order for the valley to be optically active, $u_{c\vb{k}=0}(\vb{r}')$ and $u_{v\vb{k}'=0}(\vb{r}')$ carry different representations of $C_n$, and therefore when the system rotates, the phase factors of the two factors in this term do not cancel.
From this, it follows that this term has an angular momentum of $\tau(m_c - m_v) = w$ when perceived as a matrix acting on the momentum degree of freedom (i.e. the excitonic Hilbert space).
Thus we see that the $GW$-BSE kernel, when viewed as a matrix acting on the excitonic Hilbert space, increases or decreases the angular momentum quantum number by $\pm w$.

\section{Transient absorption under coherent driving of C excitons}\label{appx:c-drive}

\begin{figure*}
    \includegraphics[width=\textwidth]{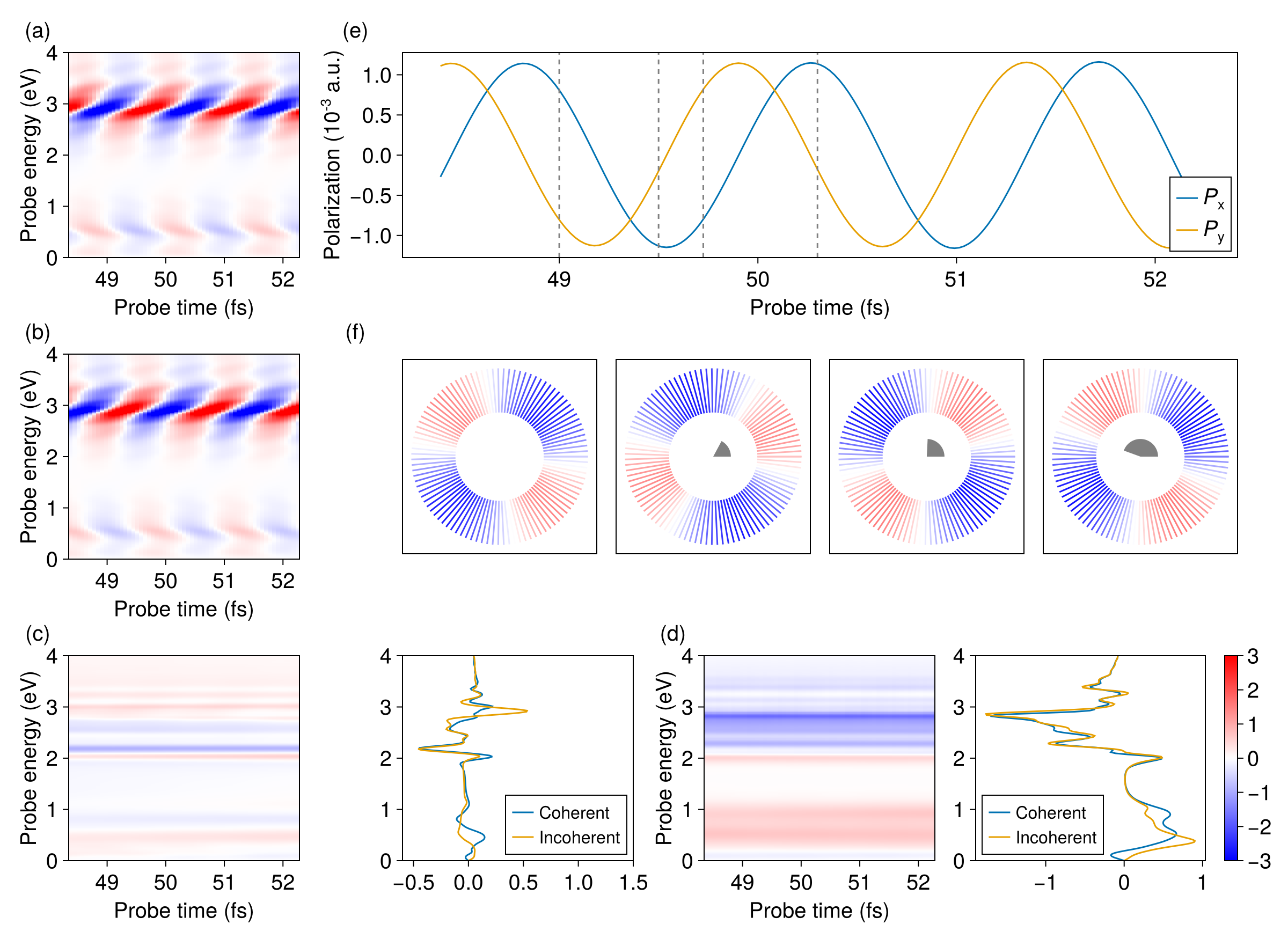}
    \caption{
    (a) Heatmap representing the imaginary parts of $\chi_{yx} + \chi_{xy}$, scaled by the same prefactor used in calculating the transient absorption spectrum as in \prettyref{fig:setup}(b).
    (b) Same as (a) but for $\chi_{xx} - \chi_{yy}$.
    (c) Same as (a) but for $\chi_{xy} - \chi_{yx}$.
    (d) Same as (a) but for $\chi_{xx} + \chi_{yy}$.
    We note that the first two oscillate with the oscillation period of the C exciton, while the remaining two do not.
    (e) The macroscopic polarization density, a replica of \prettyref{fig:tr-abs}(j).
    (c-d) The right panels compare the scaled imaginary parts of $\chi_{xy} - \chi_{yx}$ and $\chi_{xx} + \chi_{yy}$, and the corresponding values from calculations in which the non-diagonal parts of the pump-induced density matrix change $\var{\rho^{\text{pump}}}$ are zeroed out, on which linear probing is performed.
    The two sets of curves largely agree with each other, which means that the stationary part of non-equilibrium renormalization of the absorption spectrum can largely be explained by incoherent effects of the exciton population created by the pump.
    (f) Angle-dependence of the non-equilibrium renormalization of the transient absorption spectrum, measured at \SI{2.91}{eV}.
    The four panels correspond to the four vertical dashed lines in (e).
    Because the first-order influence of the coherent exciton population to the transient absorption spectrum is restricted to the $m = \pm 2$ part, the non-equilibrium renormalization of the transient absorption spectrum takes the form of $A(t) \cos 2 \theta + B(t) \sin 2 \theta$, in which the coefficients oscillate sinusoidally with the frequency of the C exciton, meaning that in a single period of coherent C oscillation, the angle dependence of transient absorption rotates \SI{180}{\degree}.}
    \label{fig:oscillate}
\end{figure*}

\prettyref{fig:oscillate} illustrates the transient absorption spectrum in monolayer \ce{MoS2} under coherent dressing of C excitons,
in which we observe the rotating $\cos 2\theta$ and $\sin 2\theta$ optical anisotropy and the delay time dependence of the four symmetric sections of the   electric susceptibility tensor.
\bibliography{tdagw,floquet,experiment,other-states,mos2}

\end{document}